\documentclass[aps,pra,manuscript,superscriptaddress,showpacs]{revtex4-1}
\usepackage{amsmath}
\usepackage{amssymb}
\usepackage{graphicx}
\usepackage{verbatim}
\usepackage{xcolor}
\usepackage{epstopdf}

\begin{document}

\title{N$_2^+$ Lasing: Gain and Absorption in the Presence of Rotational Coherence}

\author{Marianna Lytova}
\email[]{mlytova2@uottawa.ca}
\affiliation{National Research Council of Canada, 100 Sussex Drive, Ottawa ON K1A 0R6, Canada}
\affiliation{Department of Physics, University of Ottawa, Ottawa, Canada, K1N 6N5}

\author{Maria Richter} 
\affiliation{Max-Born Institute, Max-Born-Strasse 2A, D-12489, Berlin, Germany}

\author{Felipe Morales} 
\affiliation{Max-Born Institute, Max-Born-Strasse 2A, D-12489, Berlin, Germany}

\author{Olga Smirnova}
\affiliation{Max-Born Institute, Max-Born-Strasse 2A, D-12489, Berlin, Germany}
\affiliation{Technische Universit\"at Berlin, Ernst-Ruska-Geb\"aude, Hardenbergstra\ss e 36A, 10623 Berlin, Germany}

\author{Misha Ivanov}
\affiliation{Max-Born Institute, Max-Born-Strasse 2A, D-12489, Berlin, Germany}
\affiliation{Blackett Laboratory, Imperial College London,
SW7 2AZ London, United Kingdom}
\affiliation{Department of Physics, Humboldt University, Newtonstra\ss e 15, D-12489, Berlin, Germany}

%\thanks{}
\author{Michael Spanner}
\email[]{michael.spanner@nrc.ca}
\affiliation{National Research Council of Canada, 100 Sussex Drive, Ottawa ON K1A 0R6, Canada}
\affiliation{Department of Physics, University of Ottawa, Ottawa, Canada, K1N 6N5}

\date{\today}

\begin{abstract}
We simulate the pump-probe experiments of lasing in molecular nitrogen ions
with particular interest on the effects of rotational wave-packet dynamics. Our
computations demonstrate that the coherent preparation of rotational wave
packets in N$_2^+$ by an intense short non-resonant pulse results in a
modulation of the subsequent emission from $B^2\Sigma_u^+ \rightarrow
X^2\Sigma_g^+$ transitions induced by a resonant seed pulse.  We model the
dynamics of such pumping and emission using density matrix theory to describe
the N$_2^+$ dynamics and the Maxwell wave equation to model the seed pulse
propagation.  We show that the gain and absorption of a delayed seed pulse is
dependent on the pump-seed delay, that is, the rotational coherences excited by
the pump pulse can modulate the gain and absorption of the delayed seed pulse.
Further, we demonstrate that the coherent rotational dynamics of the nitrogen
ions can cause lasing without electronic inversion. 
\end{abstract}

%\pacs{33.20.Sn, 33.80.−b, 33.80.Be}
%\keywords{}

\maketitle

%%%%%%%%%%%%%%%%%%%%%%%%%%%%%%%%%%%%%%%%%%%%%%%%%%%%%%%%%%%%%%%%%%%%%%%%%%%%%%%%%
%%%%%%%%%%%%%%%%%%%%%%%%%%%%%%%%%%%%%%%%%%%%%%%%%%%%%%%%%%%%%%%%%%%%%%%%%%%%%%%%%
\section{Introduction}

Laser-induced molecular alignment of polarizable molecules was first considered
by Friedrich and Herschbach \cite{Friedrich95}. This initial work was followed
by numerous experimental and theoretical studies that further developed the
scope of laser-induced molecular alignment \cite{Larsen99, Stapel03, Dooley03}.
An important result was the realization that ultrashort pulses could be used to
generate time-dependent alignment that persists after the initial laser pulse
has past \cite{Seideman99,Ortigoso99,Rosca01}, a phenomenon called rotational
wave-packet revivals \cite{Eberly, Averb}.  The rotational revivals can in turn
be used to shape and control light pulses propagating through the rotationally
excited medium, e.g., for pulse compression down to the single-cycle limit
\cite{Kalosha02,Bartels02} through spectral broadening and shaping, or
creating quantum optical memory \cite{Thekk16} by forcing absorption and
emission of light.  In this paper, we explore the effects of rotational wave
packets on the absorption and amplification of a delayed seed pulse propagating
through a rotationally-excited gas of N$_2$/N$_2^+$ molecules.  Such a system
arises naturally in the so-called air laser, a phenomenon that was originally
observed in 2003 in laser filaments driven by ultrashort strong laser pulses
propagating in air \cite{Chin03}, and has recently become subject of numerous
advanced experimental studies \cite{Yao11, Liu13, Ni13, Zhang13, Zeng14, Xu15,
Yao16, Azarm, Arissian18, Britton18, Britton19, Kartashov14}.

Following the initial observations, the lasing process was generalized to a
pump-probe scheme where a strong pump pulse ionizes and excites N$_2$ gas
followed by a seed pulse that is amplified at selected wavelengths
\cite{Yao11,Liu13}.  More specifically, the seeded process proceeds as follows
First, the leading edge of the pump pulse exerts a torque on the neutral
nitrogen molecules towards the laser polarization direction, preparing coherent
rotational wave packets in N$_2$. Near the peak of the pump, a fraction of the
rotationally excited molecules is strong-field ionized, producing  rotationally
excited N$_2^+$ ions in the ground ($X^2\Sigma_g^+$) and lowest excited
($A^2\Pi_u$, $B^2\Sigma_u^+$) electronic levels that are driven further during
the trailing edge of the pulse. The weak seed pulse that follows is tuned to
the $B^2\Sigma_u^+ \leftrightarrow X^2\Sigma_g^+$ transition energy in N$_2^+$
(391 nm).  It is seen to undergo exponential gain \cite{Yao11}, which, with
varying seed delay, is modulated by the long-lived rotational dynamics of the
ions induced during the pump step \cite{Zhang13}.  The observed gain of the
time-delayed seed is widely held as evidence that the pump pulse creates
population inversion between the $B^2\Sigma_u^+$ and $X^2\Sigma_g^+$ states in
the ion \cite{Ni13}, but also inversionless mechanisms \cite{Mysyrowicz} have
been proposed to cause the amplification.  

In this paper, we explore the process of absorption and gain of the seed pulse
in the presence of coherent rotational excitations, which is the transient
inversion induced by the rotational wave packets evolving on the
$X^2\Sigma_g^+$ and $B^2\Sigma_u^+$ surfaces proposed in
Ref.\cite{Kartashov14}.  This process is effectively a manifestation of
ultrafast lasing without inversion scenarios \cite{OlgaK1,OlgaK2,OlgaK3}.  
There are two key ingredients that make N$_2^+$ lasing possible without
inversion. First, the parallel coupling between the $X^2\Sigma_g^+$ and
$B^2\Sigma_u^+$ states; molecules that are aligned with the seed polarization
absorb and emit more efficiently than those perpendicular to it. Second, the
different rotational constants of the ionic states that lead to a temporal
offset in their rotational evolutions.         
There will be moments in time where the molecules in the $B^2\Sigma_u^+$ state
are preferentially aligned with the seed polarization while those in the
$X^2\Sigma_g^+$ are preferentially aligned perpendicular, thus giving an
advantage to the emission from $B^2\Sigma_u^+$ over the absorption from
$X^2\Sigma_g^+$ even in the absence of electronic inversion.  Thus, the lasing
regime can be achieved due to rotational wave packet evolution of the excited
and ground states, even if there is no explicit electronic population inversion.
Based on this idea, the authors of Ref.\cite{Kartashov14} proposed the
condition of lasing in the form
$p_B\langle\cos^2\theta\rangle_B(t)>p_X\langle\cos^2\theta\rangle_X(t)$, i.e.
the product of the population and  the molecular alignment measure for the
excited state $B^2\Sigma_u^+$ is higher than the same parameter in the ground
state $X^2\Sigma_g^+$. Here, we develop computational models that confirm the
role of the rotational transient inversion mechanism \cite{Kartashov14}.

%%%%%%%%%%%%%%%%%%%%%%%%%%%%%%%%%%%%%%%%%%%%%%%%%%%%%%%%%%%%%%%%%%%%%%%%%%%%%%%%%
%%%%%%%%%%%%%%%%%%%%%%%%%%%%%%%%%%%%%%%%%%%%%%%%%%%%%%%%%%%%%%%%%%%%%%%%%%%%%%%%%
\section{Computational model}\label{full_mod}

Our interest is to investigate the gain/absorption process of a delayed seed in
the presence of the rotationally-excited $X^2\Sigma_g^+$ and $B^2\Sigma_u^+$
ionic states. To this end, we start by modeling the generation of the
rotational wave packets by the pump pulse.  Both the initial rotational
excitation in the neutral and subsequent rotational excitation in the ion
states following ionization are computed.  Focusing primarily on the effects
of the rotational coherences on the gain process, we do not attempt to fully
model the possible inversion generated by the pump pulse in the present study,
but take the liberty to vary the relative populations of the ionic states
directly to seeing how the rotational coherences can affect the gain/absorption
process in both inverted and non-inverted scenarios.  With this goal in mind,
we also omit the inclusion of the $A^2\Pi_u$ electronic state in N$_2^+$ which
is known to cause depletion of the $X^2\Sigma_g^+$ state population through a
one-photon coupling \cite{Xu15,Yao16}, since we are instead choosing to set the
$X^2\Sigma_g^+$ and $B^2\Sigma_u^+$ populations by hands.  Our 
treatment of rotational lasing without inversion in N$_2^+$ that includes a
more complete modeling of the pump pulse excitation is presented elsewhere
\cite{MBI_airlasing_paper}.  After the pump pulse has generated the
rotationally-excited medium, we then solve the coupled Maxwell/von Neumann
equations for the propagation of the seed pulse through the excited medium in
order to calculate the gain and/or absorption of the seed.  

We choose to model the quantum system using a density matrix approach.  First,
in the case where there is an initial thermal distribution, we have found the
density matrix approach to be computationally faster than using the
Schr\"odinger equation; the latter requires averaging over separate simulations
for each rotational state in the initial ensemble, while the former can group
many incoherently populated initial rotational states into a single simulation.
Second, using a density matrix approach allows us to naturally incorporate the
case where the $X^2\Sigma_g^+$ and $B^2\Sigma_u^+$ states of the ion have no
initial coherence relative to each other, which is necessary to model the
physical situation where the system does not emit unless the emission is
triggered by a seed. The following outlines the details of our model for this
study.

%%%%%%%%%%%%%%%%%%%%%%%%%%%%%%%%%%%%%%%%%%%%%%%%%%%%%%%%%%%%%%%%%%%%%%%%%%%%%%%%
\subsection{Level Structure and Initial Conditions}\label{SecInitialConditions}

\begin{figure}[t]
    \includegraphics[width=0.75\columnwidth]{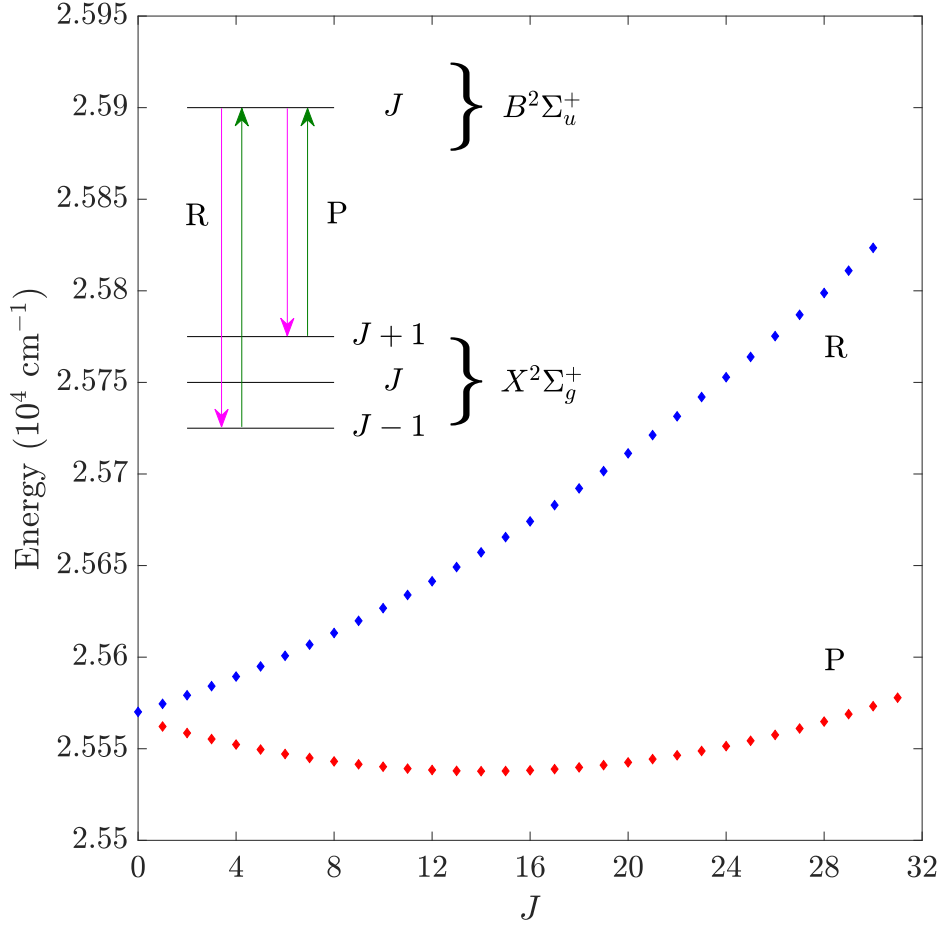}
	\caption{\label{branches} Energy diagram of the P- and R-branches for radiative
	transitions $B^2\Sigma_u^+ \leftrightarrow X^2\Sigma_g^+$ in the vibrational
	ground state $v=0$.}
\end{figure}

In the remainder of the paper, we refer to the neutral state $X^1\Sigma_g^+$ and
the ionic states $X^2\Sigma_g^+$ and $B^2\Sigma_u^+$ as $N$, $X$ and $B$. Their energy
levels are written in standard form \cite{Klyn82}:
\begin{eqnarray}\label{Etot}
	\nonumber \mathcal{E}_{J}&=&T_e+(B_e-\tfrac{\beta_e}{2}) J(J+1)-D_eJ^2(J+1)^2\\
	&+&\omega_e\big(\nu+\tfrac{1}{2}\big)-X_e\omega_e\big(\nu+\tfrac{1}{2}\big)^2+Y_e\omega_e\big(\nu+\tfrac{1}{2}\big)^3,
\end{eqnarray}
where the rotational and vibrational constants are summarized in Table
\ref{rot_const}. We designate the ro-vibronic energies of $N$, $X$ and $B$  as
$E_J^N$, $E_J^X$ and $E_J^B$, respectively. The minimum electronic energies,
$T_e$, of $N$ and $X$ are set to zero, while for $B$ we use $T_e$=25461.4
$cm^{-1}$. For the neutral, we neglect the vibrational corrections to the
rotational energies. The energies of the P- and R-branches are presented in
Fig.~\ref{branches}, with the inset illustrating the definition of both
branches.  We include only the ground vibrational state ($\nu = 0$) of each
electronic state in in the current model, which corresponds to the 391 nm
transition in $N_2^+$.

\begin{table*}[t]
\caption{\label{rot_const} Molecular constants in cm$^{-1}$ for the neutral ($X^1\Sigma_g^+$) 
and ionic states ($X^2\Sigma_g^+$ and $B^2\Sigma_u^+$) \cite{Herzberg, Klyn82}.}
\begin{ruledtabular}
\begin{tabular}{lllllll}
State                   & $B_e$    & $D_e$               & $\beta_e$ & $\omega_e$    & $X_e\omega_e$ & $Y_e\omega_e$    \\ \hline
$X^1\Sigma_g^+$ (N) & 1.989581 & $5.76\cdot 10^{-6}$ & (0)       & (0)           & (0)           & (0)              \\
$X^2\Sigma_g^+$ (X) & 1.93176  & $6.1\cdot 10^{-6}$  & 0.01881   & 2207.00       & 16.10         & -0.040           \\
$B^2\Sigma_u^+$ (B) & 2.07456  & $6.17\cdot 10^{-6}$ & 0.024     & 2419.84       & 23.18         & -0.537           \\
\end{tabular}
\end{ruledtabular}
\end{table*}

The initial N$_2$ medium is taken to be at room temperature ($T=298$ K)
with number density $N_{mol} = 5 \times 10^{18}$ cm$^{-3}$.  The rotational
levels of the neutral are incoherently populated according to the Boltzmann
distribution 
\begin{equation}
	{\cal P}_{\cal B}(J,M)=\dfrac{g_Je^{-\mathcal{E}^{N}_J/kT}}{\sum_J g_J(2J+1) e^{-\mathcal{E}^{N}_J/kT}} ,
\end{equation}
where $J$ is the total angular momentum of a particular N$_2$ molecule in the
initial ensemble, and $M$ is the projection of the angular momentum onto 
the z-axis of the laboratory frame, which is aligned along the polarization
direction of the pump and the seed.  The factor $g_J$ takes into account the nuclear spin statistics
\cite{Herzberg}, which for N$_2$ \cite{Dooley03} is
\begin{equation}\label{NucSpin}
 g_J=\left\{\begin{array}{l}
        2, \text{ for even } J\\
        1, \text{ for odd } J.
       \end{array}\right.
\end{equation}
Note that in the initial thermal ensemble, the population of each $J$ state is
evenly distributed across its  $M=0,+-1, ..., +-J$ sublevels. 

Using linearly polarized pump and seed pulses, $M$ is conserved throughout the
dynamics due to cylindrical symmetry about the laser polarization direction.
We therefore build a set of $(J,J')$-dependent density matrices, with one matrix
for each group of initial rotational states with common quantum number $|M|$. The
quantum dynamics are then computed separately for each $|M|$-subset of states. In
the following, we omit the $M$ dependence of the density matrices, but wherever
the summation over $M$ is required, we will make it explicit in the equations.

The form that we adopt for the density matrix $\hat\rho^T(t)$ for the total
system is
\begin{equation}\label{EqRhoN}
	\hat\rho^T(t) = \hat\rho^N(t) + \hat\rho^I(t),
\end{equation}
where 
\begin{equation}\label{EqRhoN}
	\hat\rho^N(t) = \sum_{JJ'} |N\rangle |JM\rangle \rho^N_{JJ'}(t) \langle J'M|\langle N|
\end{equation}
is the density matrix for the neutral, and
\begin{equation}\label{EqRhoI}
 \hat\rho^I(t)=\left[ {\begin{array}{cc}
                         \hat\rho^X(t) & \hat\rho^{XB}(t) \\
                          \hat\rho^{BX}(t) & \hat\rho^B (t)
                      \end{array}}\right],
\end{equation}
is the density matrix of the ion that has been further split into the density
matrices for the $X$ and $B$ states
\begin{subequations}
\begin{eqnarray}\label{EqRhoXBdiag}
	\hat\rho^X(t) &=& \sum_{JJ'} |X\rangle |JM\rangle \rho^X_{JJ'}(t) \langle J'M|\langle X|,  \\
	\hat\rho^B(t) &=& \sum_{JJ'} |B\rangle |JM\rangle \rho^B_{JJ'}(t) \langle J'M|\langle B|,
\end{eqnarray}
\end{subequations}
and
\begin{subequations} \label{EqRhoXBcoherence}
\begin{eqnarray}
	\hat\rho^{XB}(t) &=& \sum_{JJ'} |X\rangle |JM\rangle \rho^{XB}_{JJ'}(t) \langle J'M|\langle B|,  \\
	\hat\rho^{BX}(t) &=& \sum_{JJ'} |B\rangle |JM\rangle \rho^{BX}_{JJ'}(t) \langle J'M|\langle X| 
\end{eqnarray}
\end{subequations}
represent the coherences between the $X$ and $B$ electronic states.
In Eqs.~(\ref{EqRhoN}) to \eqref{EqRhoXBcoherence},
$|JM\rangle$ are the standard spherical harmonic rotational
functions $\langle \theta,\varphi|JM\rangle$ = Y$_J^M(\theta,\varphi)$ with 
$\theta=0$ corresponding to the same $z$-axis along which the laser pulses are
polarized, $|N\rangle$ represents the electronic state of the neutral, and
$|X\rangle$ and $|B\rangle$ represent the electronic states of the ion.  With
these definitions, the initial conditions at $t=0$ become
\begin{subequations}\label{EqInitialRho}
\begin{equation}\label{EqInitialRhoN}
	\rho^N_{JJ'}(0) = {\cal P}_{\cal B}(J,M) \: \delta_{JJ'} \: u[J-|M|] \\ 
\end{equation}
and
\begin{equation}\label{EqInitialRhoXB}
	\rho^X_{JJ'}(0) = \rho^B_{JJ'}(0) = \rho^{XB}_{JJ'}(0) = \rho^{BX}_{JJ'}(0) = 0, 
\end{equation}
\end{subequations}
where $\delta$ is the Kronecker delta, and $u$ is the Heaviside discrete step function.

%%%%%%%%%%%%%%%%%%%%%%%%%%%%%%%%%%%%%%%%%%%%%%%%%%%%%%%%%%%%%%%%%%%%%%%%%%%%%%%%
\subsection{Pump stage}

The pump laser pulse 
is defined as ${\bf E}_p(t) = {\boldsymbol \epsilon}_{z} F_p(t)
\cos(\omega_p t)$, with the envelope given by
\begin{equation}
F_p(t)=E_{p0}\left\{\begin{array}{ll}
             \sin(\pi t/\tau_{on}), & \quad 0\leq t< \tau_{on}\\
             0, & \quad t\geq\tau_{on}
            \end{array}
   \right.
\end{equation}
and ${\boldsymbol \epsilon}_{z}$ is the unit vector pointing along the $z$
direction.  This choice of $F_p(t)$ gives a ``$\sin^2$" envelope for the
intensity of the pump, where the full width at half maximum (FWHM) of the
intensity profile is given by FWHM = $\tau_{on}/2$.  We make the assumption
that the pump pulse does not undergo significant change as it propagates
through the medium, and hence there is no dependence of the equations used in
the pump step along the propagation direction $y$.

The time evolution of the density matrices is described
using the von Neumann equation, which in atomic units (a.u.) is
\begin{equation}\label{Liouv}
	 i \frac{\partial\hat\rho^k(t)}{\partial t}=\big[\hat H^k(t),\hat\rho^k(t)\big],
\end{equation}
where $k={N,X,B,I}$ represents the evolution equation for the neutral,
$X$ state, $B$ state, and total ionic system respectively.
The neutral Hamiltonian operator is
\begin{equation}\label{EqHamil}
	\hat H^N(t) = \sum_J \mathcal{E}^{N}_J |N\rangle|JM\rangle \langle JM|\langle N| + U(t,\theta) |N\rangle\langle N|,
\end{equation}
where the first term on the left hand side is the rotational kinetic energy,
and the second term is the interaction potential between the polarizability of
the neutral molecule and pump laser field \cite{Friedrich95,Stapel03,Boyd} with
\begin{equation}\label{potential}
	U(t,\theta)=-\dfrac{1}{4}(\alpha_\perp^N+\Delta\alpha^N\cos^2\theta)F_p^2(t).
\end{equation}
In this last equation (\ref{potential}), $\theta$ 
is the angle between the internuclear axis and the laser polarization direction,
and the
polarizability anisotropy is $\Delta\alpha^N=\alpha_\parallel^N-\alpha_\perp^N$,
where $\alpha_\parallel^N$ and $\alpha_\perp^N$ are the parallel and perpendicular
elements of the polarizability tensor of the neutral (see Table
\ref{polariz_const}).  In the $|JM\rangle$ basis, the Hamiltonian matrix for
the neutral is composed of
\begin{eqnarray}\label{Ham}
 \nonumber H^N_{JJ'}(t)&=&\langle JM|\hat H^N | J'M \rangle  \\
 && =\big(\mathcal{E}^{N}_J + U_\perp(t)\big)\delta_{JJ'} + U_0(t)R_{JJ'} ,
\end{eqnarray}
where 
\begin{subequations}
\begin{eqnarray}
	U_\perp(t)&=&-\alpha_\perp^N F_p^2(t)/4,   \\ 
	U_0(t)    &=&-\Delta\alpha^N F_p^2(t)/4,
\end{eqnarray}
\end{subequations}
and $R_{JJ'}$ are the matrix elements of $\cos^2\theta$
\begin{equation}\label{Rcoeff}
	R_{JJ'}=\langle JM|\cos^2\theta|J'M\rangle.
\end{equation}
$R_{JJ'}$ are non-zero only if $J'=\{J-2, J, J+2\}$, allowing Raman transitions
with $\Delta J = \pm 2$.  Note that the $R_{JJ'}$ formally depend on $M$. In
the following we will however omit their and all other matrix elements'
$M$-dependence for clarity.

Due to the the exponential dependence of strong-field ionization on the
instantaneous intensity of the driving laser pulse \cite{Keldysh}, we let
ionization take place at the peak of the pump pulse at time $t=\tau_{on}/2$, which
results in the population of the $X$ and $B$ ionic states.  We construct the
the ionic density matrices from the neutral density using the following steps.

First, we account for the angular dependence of the ionization probability
${\cal P}_I(\theta)$ estimating it as a ``peanut'' shape
\cite{Pavicic,Spanner13}
\begin{equation}\label{peanut}
	{\cal P}_I(\theta) = \cos^2\theta+\frac{1}{2}.
\end{equation}
During the ionization step, the density that is transferred into the ionic
states acquires this additional angular dependence.  We take this into
account by constructing an intermediate density matrix
\begin{equation}
	\hat\rho'=\mathcal{C} \, \hat Q \, \hat\rho^N(\tau_{on}/2) \, \hat Q.
\end{equation}
where the matrix elements of $\hat Q$ are defined as
\begin{eqnarray}\label{Tcoeff}
	Q_{JJ'} &=& \langle JM|{\cal P}_I(\theta)|J'M\rangle \\ \nonumber
	        &=& R_{JJ'} +\dfrac{1}{2} \delta_{JJ'} u[J-|M|],
\end{eqnarray}
with the same definitions for $\delta$ and $u$ as in Eq.~\eqref{EqInitialRho},
and the normalization factor is
\begin{equation}
	\mathcal{C}= \sum_M \text{tr}(\hat\rho^N(\tau_{on}/2)) \, / \, \sum_M \text{tr}(\hat Q\,\hat\rho^N(\tau_{on}/2)\,\hat Q),
\end{equation}
where the summation over all the possible $M$-subsets in the initial thermal
distribution is applied (recall Section \ref{SecInitialConditions} for
discussion about $M$-subsets).  The inclusion of the normalization factor
$\mathcal C$ has the effect of preserving the norm of the total density matrix before
and after the angular ionization probability ${\cal P}_I(\theta)$ is applied.  This allows
us to set the ionized fraction by using a scaling parameter $\eta$.  We
designate the fraction of N$_2^+$ ions with respect to neutrals as $\eta$,
setting this value by hand. We also allow ourselves to vary the relative
populations of the $X$ and $B$ states of the ion immediately after ionization,
denoting these relative populations as $p_X$ and $p_B$ defined such that
$p_X+p_B=1$.  Hence, $\eta\, p_X$ and $\eta\, p_B$  are the total populations
in the $X$ and $B$ states.

Second, special consideration of the nuclear spin statistics must be taken
\cite{Herzberg}.  While the electronic symmetry does not change during the
$N\rightarrow X$ ionizing transition, it does during the $N\rightarrow B$
transition.  This change in symmetry should be accompanied by a flip in the
nuclear spin statistics $g_J
\rightarrow g'_J$ where
\begin{equation}\label{NucSpin}
 g'_J=\left\{\begin{array}{l}
        1, \text{ for even } J\\
        2, \text{ for odd } J.
       \end{array}\right.
\end{equation}
In principle, if one would compute the ionization step with both the electronic and
rotational degrees of freedom included rigorously and consistently, the
only appearance of the nuclear spin factors would be in the initial Boltzmann
distribution.  For example, in one-photon ionization
where the electronic and rotational degrees of freedom can be included on the same
footing in first-order perturbation theory, the switch from $g_J$ to $g'_J$
occurs automatically without needing to account for this flip by hand.
In the case of strong-field ionization, however, a rigorous treatment 
of the ionization step that includes both the electronic and rotational
degrees of freedom is not currently available, and we must account for
the switch $g_J \rightarrow g'_J$ by hand during the $N\rightarrow B$ transition.
This is accomplished when constructing $\hat\rho^B$ by i) dividing out
the $g_J$ factor from $\hat\rho'$, ii) multiplying in the $g'_J$. 
We label the resulting intermediate density matrix $\hat \rho''$.

With these steps in hand, the initial conditions for the density matrices in
the ionic states that are populated by ionization at the peak of the pulse are
then given by
\begin{eqnarray}
	\hat\rho^X(\tau_{on}/2) &=& \eta \, p_X \, \hat \rho', \label{eqn:1}\\
	\hat\rho^B(\tau_{on}/2) &=& \eta \, (1-p_X)\, \hat \rho''. \label{eqn:2}
\end{eqnarray}
The coherences between the $X$ and $B$ states remain zero during the ionization
step
\begin{equation}
	\rho^{XB}_{JJ'}(\tau_{on}/2) = \rho^{BX}_{JJ'}(\tau_{on}/2) = 0.
\end{equation}
This is appropriate because, as mentioned above, we are considering
the case where the emission must be seeded, which implies that
there is no electronic coherence generated in the ion following 
ionization.  From a physical point of view, the lack of coherence
in the ion is due to the fact that the liberated electron is 
entangled with the ionic core, and tracing out the continuum
electron degree of freedom decoheres the $X$ and $B$ ionic states.

For times $t \geq \tau_{on}/2$, $\hat\rho^X(t)$ and $\hat\rho^B(t)$ continue to
evolve under the influence of the second half of the pump pulse and undergo
further rotational excitation.  This additional rotational excitation is
included by solving the von Neumann equation \eqref{Liouv} for the propagation
of the coefficients $\rho^X_{JJ'}(t)$ and $\rho^B_{JJ'}(t)$.  The $\hat
H^X(t)$ and $\hat H^B(t)$ Hamiltonians used when solving Eq.~\eqref{Liouv} for
the propagation of $\hat\rho^X(t)$ and $\hat\rho^B(t)$ have the 
analogous form to Eq.~\eqref{Ham}
\begin{subequations}
\label{EqHamilXandB}
\begin{eqnarray}
	\hat H^X(t) &=& \sum_J \mathcal{E}^{X}_J |X\rangle|JM\rangle \langle JM|\langle X| + U(t,\theta) |X\rangle\langle X|, \\
	\hat H^B(t) &=& \sum_J \mathcal{E}^{B}_J |B\rangle|JM\rangle \langle JM|\langle B| + U(t,\theta) |B\rangle\langle B|,
\end{eqnarray}
\end{subequations}
but with $U(t,\theta)$ now using the polarizabilities corresponding to the $X$
and $B$ states, see Table~\ref{polariz_const}.  

\begin{table}[t!]
\caption{\label{polariz_const} Polarizability coefficients
calculated using the GAMESS electronic structure package \cite{GAMESS},
with the aug-cc-pVTZ basis set at a CAS MCSCF level of theory, evaluated at the
equilibrium bondlength of the neutral.}
\begin{ruledtabular}
\begin{tabular}{llll}
State & $\Delta\alpha^k$ (a.u.) & $\alpha_\perp^k$ (a.u.) \\\hline
$X^1\Sigma_g^+ \:\:(k=N)$ & 4.349 & 9.252  \\
$X^2\Sigma_g^+ \:\:(k=X)$ & 9.695 & 8.509\\
$B^2\Sigma_u^+ \:\:(k=B)$ & -4.68 & 6.582 \\
\end{tabular}
\end{ruledtabular}
\end{table}

After the pump is over, we continue the time evolution
of the density matrices up until the seed pulse arrives 
using analytical solutions that have the same form for all three components
\begin{equation}\label{field_free}
	\hat\rho^k(t)=\hat\rho^k(\tau_{on})\circ \Omega^k(t), \quad t>\tau_{on}
\end{equation}
for $k={N,X,B}$, where the symbol $\circ$ denotes the Hadamard product
(element-wise matrix multiplication), and $\Omega^k(t)$ has the matrix elements
\begin{equation}
	\Omega^k_{JJ'}(t)=\exp\big[ i(\mathcal{E}^{k}_{J'}-\mathcal{E}^{k}_{J}) (t-\tau_{on}) \big]
\end{equation}
that depend on the differences of the energies in the corresponding electronic state.

Knowing the density matrices $\hat\rho^N(t)$, $\hat\rho^X(t)$ and $\hat\rho^B(t)$ then allows us to
compute the alignment measures for all three components $k={\{N,X,B\}}$ 
\begin{equation}
	\langle\cos^2\theta\rangle^k(t)=\frac{1}{p_k}\sum_M\text{tr}(\hat \rho^k(t) \hat R),
\end{equation}
the $\hat R$ operator has the matrix elements
given in Eq.~\eqref{Rcoeff}, and $p_N = 1$ since there is only one electronic state
in the neutral that holds population.  The
$\langle\cos^2\theta\rangle^k(t)$ quantities are commonly-used observables in
the molecular alignment literature that allow us to follow the rotational
wave-packet dynamics generated in the neutral and ion.  In addition, these
quantities will be used to construct the condition for gain outlined initially
in Ref. \cite{Kartashov14}.

The coherent rotations of the molecules generated by the pump pulse
cause a time-dependent refractive index $n(t)$ defined by
\cite{Boyd}
\begin{eqnarray}\label{refract}
	n^2(t)&=&1+4\pi N_{mol}\Big[(1-\eta)\alpha_\perp^N \nonumber \\
	        &&+(1-\eta)\Delta \alpha^N\langle\cos^2\theta\rangle^N(t)\Big] \nonumber \\
		&\equiv& 1+4\pi N_{mol}\Theta_1(t). 
\end{eqnarray}
In principle there should be additional terms in Eq.~\eqref{refract} due to the
rotational excitations in the ionic states, but these will have negligible
contribution to $n(t)$ relative to the neutral terms as the
fraction of ionized molecules $\eta$ is assumed to be small.
Eq.~\eqref{refract} is used below when propagating the seed pulse through the
rotationally-excited medium.

%%%%%%%%%%%%%%%%%%%%%%%%%%%%%%%%%%%%%%%%%%%%%%%%%%%%%%%%%%%%%%%%%%%%%%%%%%%%%%%%
\subsection{Seed propagation}

Our seed pulse is polarized along the $z$ direction as the pump pulse, and is
taken to propagate along the $y$ direction through the medium.  The initial
seed pulse at the start of the medium ($y=0$) is defined as
\begin{equation}
	{\bf E}_s(t,y=0) = {\boldsymbol \epsilon}_{z} E_s(t,y=0),
\end{equation}
with
\begin{equation}\label{seed}
	 E_s(t,y=0) = E_{s0} \, e^{ -4 \log 2 \left(\frac{t-t_{del}}{\sigma_s}\right)^2} \cos(\omega_s(t-t_{del})),
\end{equation}
where $\sigma_s$ is the full width at the half-maximum of the seed envelope, 
$\omega_s$ is the central frequency of the seed, and
$t_{del}$ is the delay time of the seed pulse.

Qualitatively, the time evolution of $E_s(t,y)$ proceeds as follows.  With
the seed pulse defined for all time at the entrance of the medium by
Eq.~\eqref{seed}, we first compute the response of the medium by using
$E_s(t,y=0)$ in a von Neumann equation for $\hat\rho^I(t,y=0)$ that couples the $X$
and $B$ states through the dipole interaction, using the $\hat\rho^X(t)$ and
$\hat\rho^B(t)$ computed in the pump section as initial conditions for
$\hat\rho^I(t,y=0)$. Second, once $\hat\rho^I(t,y=0)$ is calculated following the
interaction with the seed pulse, we use this $\hat\rho^I(t,y=0)$ to compute the
polarization of the medium which is then used as input into the Maxwell wave
equation to propagate $E_s(t,y=0)$ to the next spatial point along $y$. These
two steps are then repeated to continue propagating the seed pulse through the
medium, with $E_s(t,y=0)$ replaced with $E_s(t,y)$ at the current $y$ position.  
We now outline the equations used in these two steps to compute
$\hat\rho^I(t,y)$ and to propagate $E_s(t,y)$ along the $y$-direction.

The initial ionic density matrix for any $y$-coordinate is constructed as
\begin{equation}
 \hat\rho^I(t=t_s,y=0)=\left[ {\begin{array}{cc}
                         \hat\rho^X(t_s) & \O{}\\
                          \O{} & \hat\rho^B (t_s)
                        \end{array}}\right],
\end{equation}
and where $\O{}$ is a zero matrix.  The time $t=t_s$ is some point in time
after the pump pulse is over where we wish to start the time evolution of the
seed pulse.  The evolution of $\hat \rho^I(t,y=0)$ for time $t>t_s$ is carried
out by solving the von Neumann equation~\eqref{Liouv}.  The ionic Hamiltonian
$\hat H^I(t,y)$ is written as
\begin{equation}\label{EqHamilIonBlock}
 \hat H^I(t,y)=\left[ {\begin{array}{cc}
                       \hat H^{X}  & \hat H^{XB}(t,y)\\
                       \hat H^{BX}(t,y) & \hat H^{B}
                      \end{array}}\right],
\end{equation}
where
\begin{equation}\label{H0_V}
	\hat H^{XB}(t,y) = (\hat H^{BX}(t,y))^\dagger = - {\boldsymbol\mu}\cdot{\bf E}_s(t,y)
\end{equation}
accounts for the interaction between
the weak resonant seed pulse and the ionic states.  Because the seed is assumed
to be in the weak-field limit, the off-resonant polarizability interaction
analogous to $U(t,\theta)$ from Eq.~\eqref{potential} that would be induced by
the seed pulse is now negligible, so that Eqs.~\eqref{EqHamilXandB} become
\begin{subequations}
\begin{eqnarray}
	\hat H^X &=& \sum_J \mathcal{E}_J^{X} |X\rangle |JM\rangle \langle JM|\langle X|, \\
	\hat H^B &=& \sum_J \mathcal{E}_J^{B} |B\rangle |JM\rangle \langle JM|\langle B|,
\end{eqnarray}
\end{subequations}
during the seed step.
The energies $\mathcal{E}_J^{X}$ and $\mathcal{E}_J^{B}$ are computed according to
Eq.~\eqref{Etot} for $v=0$ corresponding to the 391 nm transition in N$^+_2$.
The $B$-$X$ dipole coupling is a parallel transition, in
which case the dipole interaction reduces to 
\begin{equation}
	-{\boldsymbol\mu}\cdot{\bf E}_s(t,y)=-\mu_{XB}E_s(t,y)\cos\theta, 
\end{equation}
where the transition dipole $\mu_{XB}=-0.74$ a.u. was computed with GAMESS
using the same level of electronic structure used to compute the
polarizabilities above. $\hat H^{XB}(t,y)$ can then be written as
\begin{eqnarray} \label{Ham2_1}
	\hat H^{XB}(t,y) &=&-\mu_{XB} E_s(t,y)\hat S,
\end{eqnarray}
where the matrix elements of $\hat S$ are given by
\begin{equation}\label{Scoeff}
	S_{JJ'}=\langle J M|\cos\theta|J' M\rangle.
\end{equation}
The only non-zero $S_{JJ'}$ occur when $J-J'=\pm 1$, resulting in the expected
one-photon selection rules for the transitions between rotational levels of the
$B$ and $X$ states.  Note that $\hat H^{XB}(t,y)$ will generate coherences
between the $X$ and $B$ electronic states of the ion that in turn cause absorption
and/or emission at the $X \leftrightarrow B$ transition frequencies.

After computing $\hat\rho^I (t,y)$, which describes the microscopic properties
of the medium, we can calculate the macroscopic polarization $P_\mu(t,y)$ of
the medium induced by the seed pulse along ${\boldsymbol \epsilon}_{z}$
\begin{eqnarray}\label{mu_seed}
	P_\mu(t,y)&=&\eta N_{mol} \langle {\boldsymbol \mu} \cdot {\boldsymbol \epsilon}_{z} \rangle   \\ \nonumber
	          &=&\eta N_{mol} \langle \mu_{XB} \cos\theta\rangle   \\ \nonumber
	          &=&\eta N_{mol}\, \mu_{XB}\, \sum_M \text{tr}(\hat \rho^I(t,y)\, \mathbb{S}),
\end{eqnarray}
where the transition matrix $\mathbb{S}$ has the form
\begin{equation}\label{tran_S}
	\mathbb{S}=\left[ {\begin{array}{cc}
                       \O{} & \hat S\\
                        \hat S {} & \O{}
                      \end{array}}\right].
\end{equation}
Also recall that at the pump stage, the excited rotation of the molecules 
generates a time-dependent refractive index given by Eq.~\eqref{refract}.
When the seed pulse propagates in the rotationally-excited medium, it is 
also affected by this refractive index giving rise to an additional
contribution to the macroscopic polarization of the medium seen
by the seed pulse given by
\begin{equation}\label{theta_pump}
	P_{{\mathrm N}_2}(t)=N_{mol}E_s(t,y)\Theta_1(t),
\end{equation}
where $\Theta_1(t)$ is defined by Eq.~\eqref{refract}.  
Due to the fact that we restrict ourselves to small values of 
$\eta$ (i.e. small faction of ionized molecules), $n(t)$ is effectively
the time-dependent refractive index generated by the rotationally-excited
neutral molecules, and hence we have labeled the associated polarization
in Eq.~\eqref{theta_pump} as $P_{{\mathrm N}_2}$ to indicate that this
polarization comes from the neutral N$_2$.

We compute the propagation of the electric field $E_s(t,y)$ of the seed
using a simplified Maxwell wave equation \cite{SolitonPaper}
\begin{equation}\label{transport}
	\frac{\partial E_s(t,y)}{\partial y} + \frac{1}{c}\frac{\partial E_s(t,y)}{\partial t} = -\frac{2\pi}{c}\frac{\partial P(t,y) }{\partial t},
\end{equation}
where the polarization $P(t,y)$ consists of the two terms introduced above
\begin{eqnarray}\label{EqFullP}
	P(t,y) = P_{{\mathrm N}_2}(t)+ P_\mu(t,y).
\end{eqnarray}
Eq.~\eqref{transport} is derived by including only the forward propagating waves, 
and is equivalent to the slowly-varying envelope approximation in the limit
of long pulse durations \cite{Boyd}.
We solve Eq.~\eqref{transport} in a reference frame that is moving at roughly
the average velocity of the pump pulse by introducing the new variable
$\tau=t-y/v_p$, where the velocity of this moving frame is taken to be
\begin{eqnarray}
  v_p = c\Big\{1+2\pi N_{mol}\Big[(1-\eta)\big(\alpha^N_\perp+\Delta \alpha^N/3\big)+\qquad\qquad\\
 \nonumber \eta\, p_X\big(\alpha^X_\perp+\Delta \alpha^X/3\big)+\eta\,p_B\big(\alpha^B_\perp+\Delta \alpha^B/3\big)\Big]\Big\}^{-1}.
\end{eqnarray}
Eq.~\eqref{transport} can then be written as
\begin{equation}\label{transport2}
  \frac{\partial E_s(\tau,y)}{\partial y} =\frac{1}{v_r}\frac{\partial E_s(\tau,y)}{\partial {\tau}}-\frac{2\pi}{c}\frac{\partial P(\tau,y)}{\partial {\tau}},
\end{equation}
where $v_r=cv_p/(c-v_p)$. 

The gain and/or absorption of the seed pulse is computed as the ratio of the
integrated spectral intensities after propagating in the rotationally excited
medium and the intensity before the interaction.  Specifically, we calculate
\begin{equation}\label{Gain}
	\Gamma(t_{del},z_{max})\!=\!1+\dfrac{\displaystyle\int_{\omega_P}^{\omega_R} [I_{out}(\omega,t_{del},z_{max})-I_{in}(\omega)]d\omega}
	{\displaystyle\int_{\omega_P}^{\omega_R} I_{in}(\omega)\,d\omega},
\end{equation}
where $I_{in}$ and $I_{out}$ are the spectral intensities (i.e Fourier power
spectrum) of the seed pulse at the input and output of the medium respectively,
$\omega_P$ is the minimum of the P-branch parabola (see Fig.~\ref{branches}),
and $\omega_R$ corresponds to the maximal transition in the R-branch under
consideration.   When $\Gamma > 1$ the seed has undergone gain, while 
$\Gamma < 1$ implies that absorption rather than gain of the seed has occurred.

%%%%%%%%%%%%%%%%%%%%%%%%%%%%%%%%%%%%%%%%%%%%%%%%%%%%%%%%%%%%%%%%%%%%%%%%%%%%%%%%
\subsection{Numerical Considerations}

We numerically solve the von Neumann equation Eq.~(\ref{Liouv}) using the
Runge-Kutta forth order (RK4) scheme \cite{Quart} to obtain the time evolution
of the coefficients  $\rho^N_{JJ'}(t)$  of the neutral density matrix.  For the
pump step, our RK4 propagation used a time step of $\Delta t$ = 1 fs.  The
maximum number of the rotational states was set to $J_{max}=40$ in both the
seed and pump steps, and the maximum rotational number used to average over the
initial thermal distribution of the neutral N$_2$ was $J_{max0}=30$.

Efficient propagation during the seed step requires further care due to the
disparate timescales imposed by the electronic energy separation of the $X$ and
$B$ states.  We first write the density matrix elements in terms of
slowly-varying amplitudes $A^k_{JJ'}(t)$,
\begin{subequations}
\begin{equation}
	\rho^k_{JJ'}(t)  = A^k_{JJ'}(t) e^{ i(\mathcal{E}^{k}_{J'}-\mathcal{E}^{k}_{J}) t }
\end{equation}
for $k={X,B}$, and
\begin{equation}
	\rho^{XB}_{JJ'}(t)  = A^{XB}_{JJ'}(t) e^{ i(\mathcal{E}^B_{J'}-\mathcal{E}^X_{J}) t },
\end{equation}
\begin{equation}
	\rho^{BX}_{JJ'}(t)  = A^{BX}_{JJ'}(t) e^{ i(\mathcal{E}^X_{J'}-\mathcal{E}^B_{J}) t }.
\end{equation}
\end{subequations}
Similarly, we write the electric field of the seed as
\begin{equation}
	E_s(t,y)  =  F_s(t,y) e^{i\omega_s t} + F_s^*(t,y) e^{-i\omega_s t}.
\end{equation}
This approach has the advantage of analytically incorporating the fast
oscillations related to the electronic spacing and the seed carrier wave into
the numerical propagation scheme.  We proceed to apply the RK4 method to the
von Neumann propagation of the $A^k_{JJ'}(t)$,  $A^{XB}_{JJ'}(t)$, and
$A^{BX}_{JJ'}(t)$ with a time step of 0.305 fs.  For the spatial propagation of
$F_s(t,y)$, we implement the Lax-Wendroff method leading to a 2$^{nd}$ order
numerical scheme \cite{Quart,Strik} with CFL-number $\Lambda=\Delta
y/(v_r\Delta t)=0.99$, where $\Delta y = 2.8\times 10^{-2}$ mm and $\Delta t =
0.00477$ fs are the space and time steps used in the seed pulse propagation
iterations.  The slowly-varying density matrix elements are converted from the
coarse time grid used in the von Neumann step to the fine time grid used in the
Lax-Wendroff step using spine interpolation.  To use the RK4 scheme in
propagating the ionic density matrix elements, we need to know not only
$F_s(t_n,y_k)$ and $F_s(t_{n+1},y_k)$, where $t_n$ and $t_{n+1}$ are two
neighboring time points at the k$^{th}$ spatial coordinate $y_k$, but also the
intermediate values $F_s(t_n+\Delta t/2,y_k)$. To preserve the 4$^{th}$ order
of RK4 we use the 4$^{th}$ order Lagrange approximation for these intermediate
points.

%%%%%%%%%%%%%%%%%%%%%%%%%%%%%%%%%%%%%%%%%%%%%%%%%%%%%%%%%%%%%%%%%%%%%%%%%%%%%%%%
\subsection{Perturbative Treatment of Seed-N$^+_2$ interaction}\label{sa_perturb}

In Ref.\cite{Kartashov14}, it was suggested that the gain seen in N$^+_2$
lasing should be related to the molecular alignment in $X$ and $B$. In
particular, it was proposed that the gain of the delayed seed pulse should be
proportional to the difference of the alignment in the $X$ and $B$ states at
the moment $t=t_{del}$ when the seed arrives
\begin{eqnarray}\label{Wtran}
 \nonumber W_{down \leftrightarrow up}(t_{del})=\qquad\qquad\qquad\qquad\qquad\qquad \\
 p_B\langle\cos\theta\rangle^B(t_{del})-p_X\langle\cos\theta\rangle^X(t_{del}).\qquad
\end{eqnarray}
Note that unlike $\Gamma$ defined in Eq.~(\ref{Gain}), the estimate $W_{down \leftrightarrow up}$
predicts gain when $W_{down \leftrightarrow up} > 0$, while absorption corresponds
to $W_{down \leftrightarrow up} < 0$.
In this section, we apply first-order perturbation theory to the interaction
between the weak seed pulse and the rotationally-excited medium, and
demonstrate how to recover the condition for gain in Eq.~(\ref{Wtran}).
Understanding the conditions required to recover the $ W_{down \leftrightarrow
up}$ estimate will help us to understand the cases presented below where the
fully-numerical formalism starts to diverge from this estimate.  We apply the
perturbation theory within the wavefunction formalism, which allows us to
obtain the perturbative result in the clearest way.  

The wave function for a generic rotationally-excited wave packet in the ion
after the pump pulse can be written as
\begin{eqnarray}\label{after_pump}
  |\Psi(t)\rangle&=\displaystyle\sum_J    x^{(0)}_{J } \, e^{- i\mathcal{E}_{J }^{X}t}| X\rangle | J M\rangle\\
                 &+\displaystyle\sum_{J'} b^{(0)}_{J'} \, e^{- i\mathcal{E}_{J'}^{B}t}| B\rangle | J'M\rangle,
\end{eqnarray}
where we have labeled the amplitudes of the wave function with a superscript
'(0)' to imply that they are the zeroth-order amplitudes (i.e. they do not
contain any interaction with the seed pulse.)

Consider first the process of absorption by a seed pulse that arrives at time
$t_{del}$.  Absorption physically corresponds to moving population from $X$ to
$B$, and we therefore compute the first-order corrections to the amplitudes in
$B$ that arise from seed-driven transitions from $X$ to $B$
\begin{eqnarray}\label{b_1}
 b_J^{(1)}(t_{del}) &=& i \mu_{XB}\Big[ \mathcal{F}(\omega^{BX}_{JJ-1})S_{JJ-1} x_{J-1}^{(0)}e^{ i\omega^{BX}_{JJ-1}t_{del}}  \nonumber \\
          &&+ \mathcal{F}(\omega^{BX}_{JJ+1})S_{JJ+1} x_{J+1}^{(0)} e^{ i\omega^{BX}_{JJ+1}t_{del}}\Big],
\end{eqnarray}
where $\omega^{BX}_{JJ'}=(\mathcal{E}^{B}_{J}-\mathcal{E}^{X}_{J'})$ are the
transition frequencies, $\mathcal{F}(\omega^{BX}_{JJ'})$ are the Fourier
amplitudes of the seed pulse at these frequencies, and $S_{JJ'} = \langle
JM|\cos\theta |J'M\rangle$ as was defined in Eq.~\eqref{Scoeff}.  
Since the $b^{(1)}_J$ are the excited state amplitudes generated by the seed, 
the total absorption can be estimated by summing over all the first-order $B$ populations:
\begin{widetext}
\begin{eqnarray}
	\sum_J|b^{(1)}_J(t_{del})|^2 &=&  |\mu_{XB}|^2 \sum_J \bigg[ |\mathcal{F}(\omega^{BX}_{JJ-1})|^2 \, |S_{JJ-1}|^2\, |x_{J-1}^{(0)}|^2 
	+ |\mathcal{F}(\omega^{BX}_{JJ+1})|^2 |S_{JJ+1}|^2 |x_{J+1}^{(0)} |^2 \\ \nonumber
	&&+2 \mathcal{F}(\omega^{BX}_{JJ-1})^* \mathcal{F}(\omega^{BX}_{JJ+1}) S_{JJ-1} S_{JJ+1} x_{J-1}^{(0)*}x_{J+1}^{(0)} 
	\cos([\mathcal{E}^{X}_{J+1} - \mathcal{E}^{X}_{J-1}] t_{del}) \bigg].
\end{eqnarray}
In order to eventually recover the  $W_{down \leftrightarrow up}$ estimate, we must now
make the assumption that the bandwidth is flat across all transition frequencies:
$\mathcal{F}(\omega^{BX}_{JJ'}) = constant$ for all $\omega^{BX}_{JJ'}$.  For convenience,
we choose to set $\mathcal{F}(\omega^{BX}_{JJ'}) = 1$.  This gives
\begin{eqnarray}\label{EqAlmostAtCos2}
	\sum_J|b^{(1)}_J(t_{del})|^2 &=&  |\mu_{XB}|^2 \sum_J \bigg[ (|S_{J+1,J}|^2 + |S_{J-1,J}|^2) |x_J^{(0)}|^2  \\ \nonumber
	&&+2 S_{J+1,J} S_{J+2,J+1} x_J^{(0)*}x_{J+2}^{(0)} 
	\cos([\mathcal{E}^{X}_{J+2} - \mathcal{E}^{X}_{J}] t_{del}) \bigg],
\end{eqnarray}
where we have also taken the liberty of rearranging some of the indices in the summation.
With a little algebra, Eq.~(\ref{EqAlmostAtCos2}) can be seen to be equivalent to the expression
\begin{equation}\label{EqAtCos2}
	\sum_J|b^{(1)}_J(t_{del})|^2 = p_X |\mu_{XB}|^2 \langle \cos^2\theta \rangle^X(t_{del})
\end{equation}
where $\langle ... \rangle^X(t_{del})$ means that we are taking the expectation
value of the rotational wave packet over the zeroth-order $X$ state at the time $t_{del}$,
and the appearance of the population term $p_X$ accounts for the fact that the population
in the X state is not unity.
In going from Eq.~(\ref{EqAlmostAtCos2}) to (\ref{EqAtCos2}) we have made use
of the fact that $(|S_{J+1,J}|^2 + |S_{J-1,J}|^2) = R_{JJ}$ and 
$S_{J+1,J} S_{J+1,J+2} = R_{JJ+2}$, where the $R_{JJ'}$ are the matrix elements
of $\cos^2\theta$ defined in Eq.~(\ref{Rcoeff}).  These two properties can be derived 
from the properties of spherical harmonics.

Equation (\ref{EqAtCos2}) shows that the absorption from the $X$ state is
proportional to the alignment in the $X$ state.   A corresponding expression
for the emission from the state $B$ can be analogously derived by repeating the
steps that lead from Eq.~(\ref{b_1}) to (\ref{EqAtCos2}) but now considering
the first-order corrections to the $X$ state that account for the seed-driven
transitions from $B$ to $X$.  This calculation yields
\begin{equation}\label{EqAtCos2_emission}
	\sum_J|x^{(1)}_J(t_{del})|^2 = p_B |\mu_{XB}|^2 \langle \cos^2\theta \rangle^B(t_{del}),
\end{equation}
which shows that the emission from $B$ is proportional to the alignment in the
$B$ state.  The total expected emission from the system, which would be given
by the emission from $B$ minus the absorption from $X$, can be now constructed
by combining the expressions (\ref{EqAtCos2}) and (\ref{EqAtCos2_emission}) 
\begin{eqnarray}
	\sum_J|x^{(1)}_J(t_{del})|^2 - \sum_J|b^{(1)}_J(t_{del})|^2 &=& |\mu_{XB}|^2 \bigg( p_B \langle \cos^2\theta \rangle^B(t_{del})
	- p_X \langle \cos^2\theta \rangle^X(t_{del})\bigg) \\ \nonumber
	&\equiv& |\mu_{XB}|^2 \, W_{down \leftrightarrow up}(t_{del}),
\end{eqnarray}
\end{widetext}
which gives the gain estimate Eq.~(\ref{Wtran}) proposed in
Ref.\cite{Kartashov14}.  Since $W_{down \leftrightarrow up}$ is constructed to
reflect the emission minus the absorption,  a value of $W_{down \leftrightarrow
up}>0$ predicts gain while $W_{down \leftrightarrow up}<0$ predicts absorption.
Although this result was here derived using a single wave function, the same
result is obtained using perturbation theory in the density matrix approach,
and the result still holds when averaging over an initial thermal distribution.
Further, it should be stressed that obtaining the expression for $W_{down
\leftrightarrow up}$ required that we assume a flat bandwidth.  This point will
be important below to understand cases where the gain starts to diverge from
the estimate $W_{down \leftrightarrow up}$.

%%%%%%%%%%%%%%%%%%%%%%%%%%%%%%%%%%%%%%%%%%%%%%%%%%%%%%%%%%%%%%%%%%%%%%%%%%%%%%%%%
%%%%%%%%%%%%%%%%%%%%%%%%%%%%%%%%%%%%%%%%%%%%%%%%%%%%%%%%%%%%%%%%%%%%%%%%%%%%%%%%%
\section{Results and discussion}\label{res}

\subsection{Rotational excitation and wave packet dynamics}

\begin{figure}[b]
	\includegraphics[width=0.95\columnwidth]{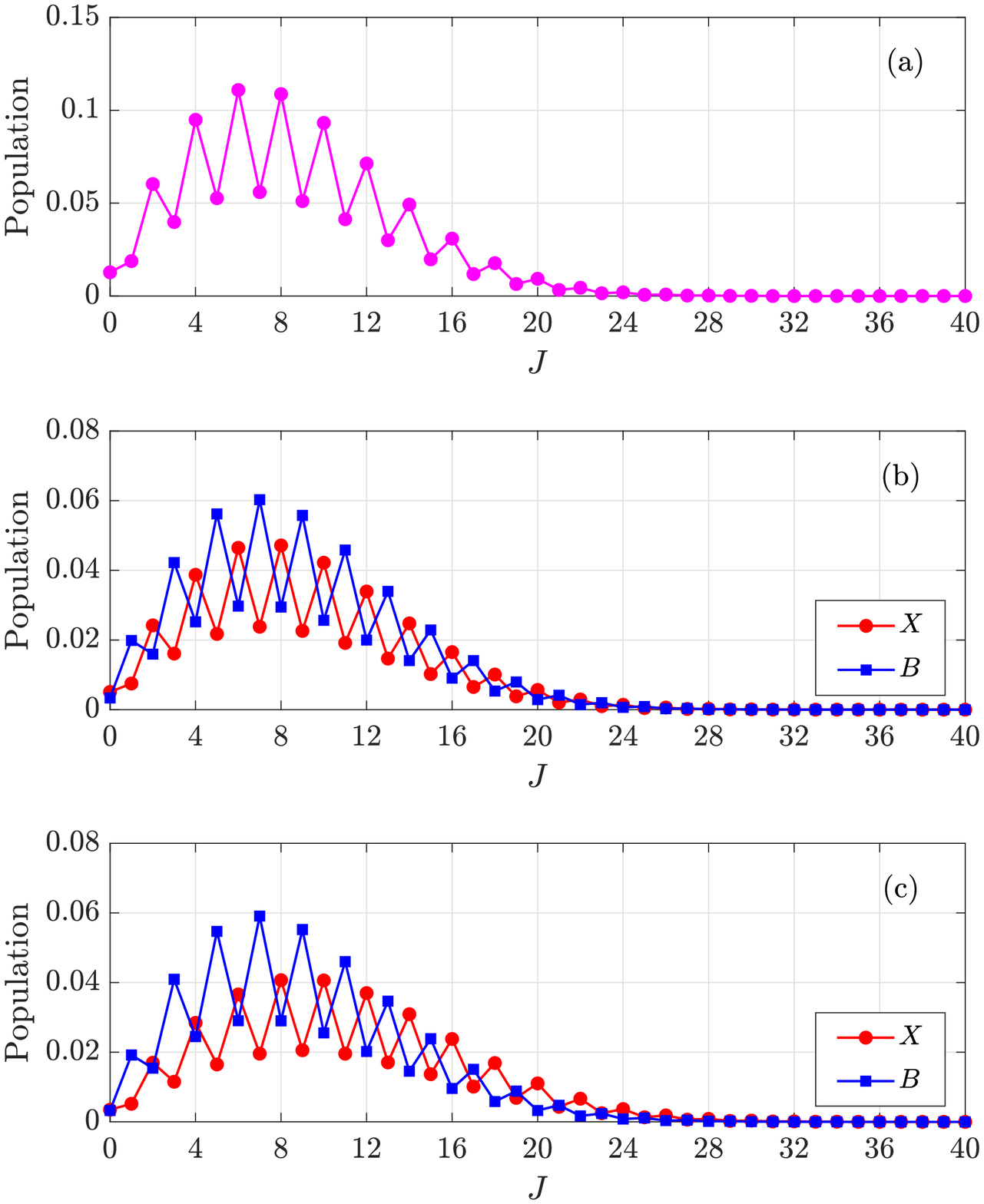}
	\caption{
	\label{pump_pop} 
	(a) Initial thermal rotational distribution 
	in the neutral with temperature $T=298$ K.  (b) Rotational populations on the  
	$X$ and $B$ ionic states after excitation/ionization by the pump pulse
	with intensity $I_{pump}=1\times 10^{14}$ W/cm$^{2}$.
	The relative ionic populations are set to be $p_X = 0.45$ and
	$p_B = 0.55$. (c) Same parameters as (b) but with $I_{pump}=2\times 10^{14}$ W/cm$^{2}$. }
\end{figure}

We first discuss the rotational excitation and rotational wave packets
generated in the pump step.  The initial thermal rotational distribution of the
neutral at temperature $T=298$ K is plotted in Fig.~\ref{pump_pop}a,
while Figs.~\ref{pump_pop}b and c show examples of the 
rotational distributions in the $X$ and $B$ ionic states after 
the pump pulse has past.
For these cases, we used a peak pump intensity
of $I_{pump}=1\times10^{14}$ W/cm$^2$ (Fig.~\ref{pump_pop}b) and
$I_{pump}=2\times10^{14}$ W/cm$^2$ (Fig.~\ref{pump_pop}c), the duration of the pump pulse
was $\tau_{on}=50$ fs (FWHM = 25 fs), and the number
density was $N_{mol}=5\times10^{18}$ cm$^{-3}$. The relative ionic populations
were set to $p_X = 0.45$ and $p_B = 0.55$.

\begin{figure}[t]
	\includegraphics[width=0.95\columnwidth]{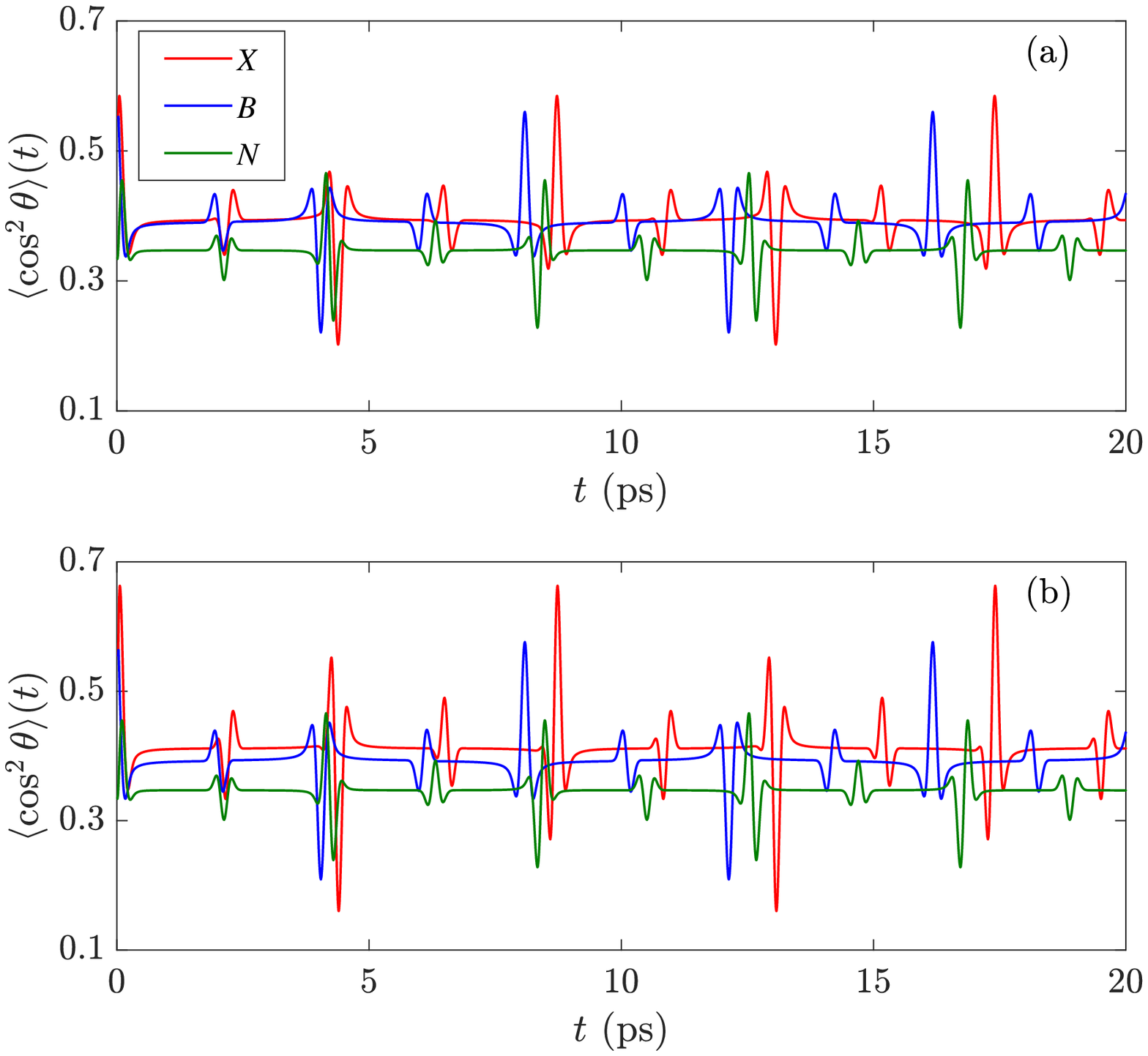}
	\caption{\label{cos2NXB} Measures of the alignment
	$\langle\cos^2\theta\rangle(t)$ for the neutral and ionic components
	at temperature $T=298$ K.  Panel (a) is for a pump intensity of 
	$I_{pump}=1\times 10^{14}$ W/cm$^{2}$, while panel (b) is
	for $I_{pump}=2\times 10^{14}$ W/cm$^{2}$.}
\end{figure}

Figures \ref{pump_pop}b and c show that both the $X$ and $B$ ionic states are
rotationally hotter than the initial neutral thermal distribution, reflecting
the rotational excitation imparted by the pump pulse.  In addition, we find that
the $X$ state is rotationally hotter than the $B$ state, an effect that is more
pronounced in the $I_{pump}=2\times10^{14}$ W/cm$^2$ case
(Fig.~\ref{pump_pop}c).  This difference between the $X$ and $B$ states is due
to the different polarizabilities for the $X$ and $B$ states, $\Delta \alpha^X$
and $\Delta \alpha^B$.  In the first half of the pump pulse, the neutral
receives a torque toward the pump polarization direction (the $z$-axis in our
case).  Following ionization, the population in $X$ continues to receive
additional torque toward the $z$-axis.  However, the population in the $B$
state receives a torque in the opposite direction since $\Delta \alpha^B$ has
the opposite sign compared to $\Delta \alpha^N$ and $\Delta \alpha^X$, and
hence the torque received on the second half of the pump pulse for the $B$ state
is partially undoing the rotational excitation imparted to the neutral on the
first half of the pump pulse.   

Figures \ref{cos2NXB}a and b plot the alignment measure
$\langle\cos^2\theta\rangle(t)$ for the neutral and ionic states after the pump
pulse, again for $I_{pump}=1\times10^{14}$ W/cm$^2$ (Fig.~\ref{cos2NXB}a) and
$I_{pump}=2\times10^{14}$ W/cm$^2$ (Fig.~\ref{cos2NXB}b), which shows the
coherent rotational dynamics that occurs following the pump pulse.
Qualitatively, a large value of  $\langle\cos^2\theta\rangle(t)$ means that the
molecules are preferentially aligned along the pump polarization direction,
while a smaller value of the alignment parameter implies that the molecules are
more aligned perpendicular to this direction.  The revivals for the different
states have different timings, which is due primarily to the different
rotational energy constant $B_e$ of each state.

%%%%%%%%%%%%%%%%%%%%%%%%%%%%%%%%%%%%%%%%%%%%%%%%%%%%%%%%%%%%%%%%%%%%%%%%%%%%
\subsection{Modulation of the seed gain}

\begin{figure}[t]
	\includegraphics[width=0.95\columnwidth]{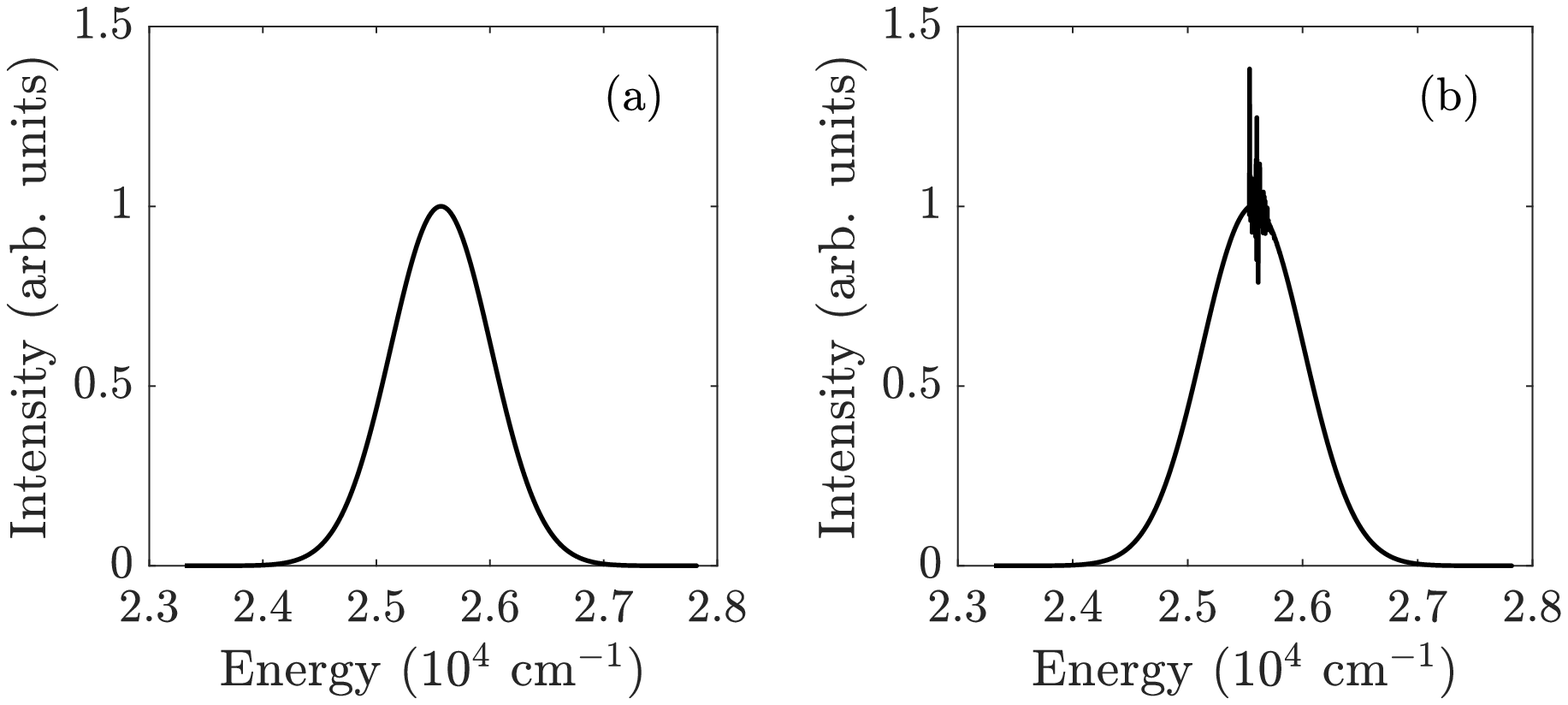}
	\caption{(a) Input spectrum of the seed pulse, $\lambda_s=391$ nm.  
	(b) Output spectrum of the seed pulse for a delay of $t_{del}$ = 5 ps 
	and total propagation length of $z_{max}$ = 0.5 mm.  Gain/absorption lines can
	been seen near the peak of the output seed spectrum. 
	}\label{FigSeedBeforeAfter}
\end{figure}

We now consider the seed propagation.  Fig.~\ref{FigSeedBeforeAfter}a
shows the spectrum of the initial seed pulse at the entrance to the medium,
while Fig.~\ref{FigSeedBeforeAfter}b shows an example of the output seed spectrum.
The input seed pulse had a peak intensity of $I_{seed} = 10^{11}$ W/cm$^2$,
a duration of $\sigma_s$ = 20 fs,
a central wavelength of $\lambda_s$ = 391 nm, and a total propagation length of
$z_{max}$ = 0.5 mm was used.  
The example case in Fig.~\ref{FigSeedBeforeAfter} 
corresponds to a delay of $t_{del}$ = 5 ps, and the relative ionic populations
were set to $p_X=0.45$ and $p_B=0.55$, and the fraction of ionized molecules
was set to $\eta$ = 0.001 (i.e. 0.1\%).   
The pump intensity used was $I_{pump}=10^{14}$ W/cm$^2$.
As can be seen in the figure, 
the output spectrum of the seed pulse has developed gain and absorption structures
in the energy region of the rotational transitions. 
The $\Gamma$ parameter in Eq.~(\ref{Gain}) is computed by integrating across
this gain/absorption window.

\begin{figure}[t]
	\includegraphics[width=0.95\columnwidth]{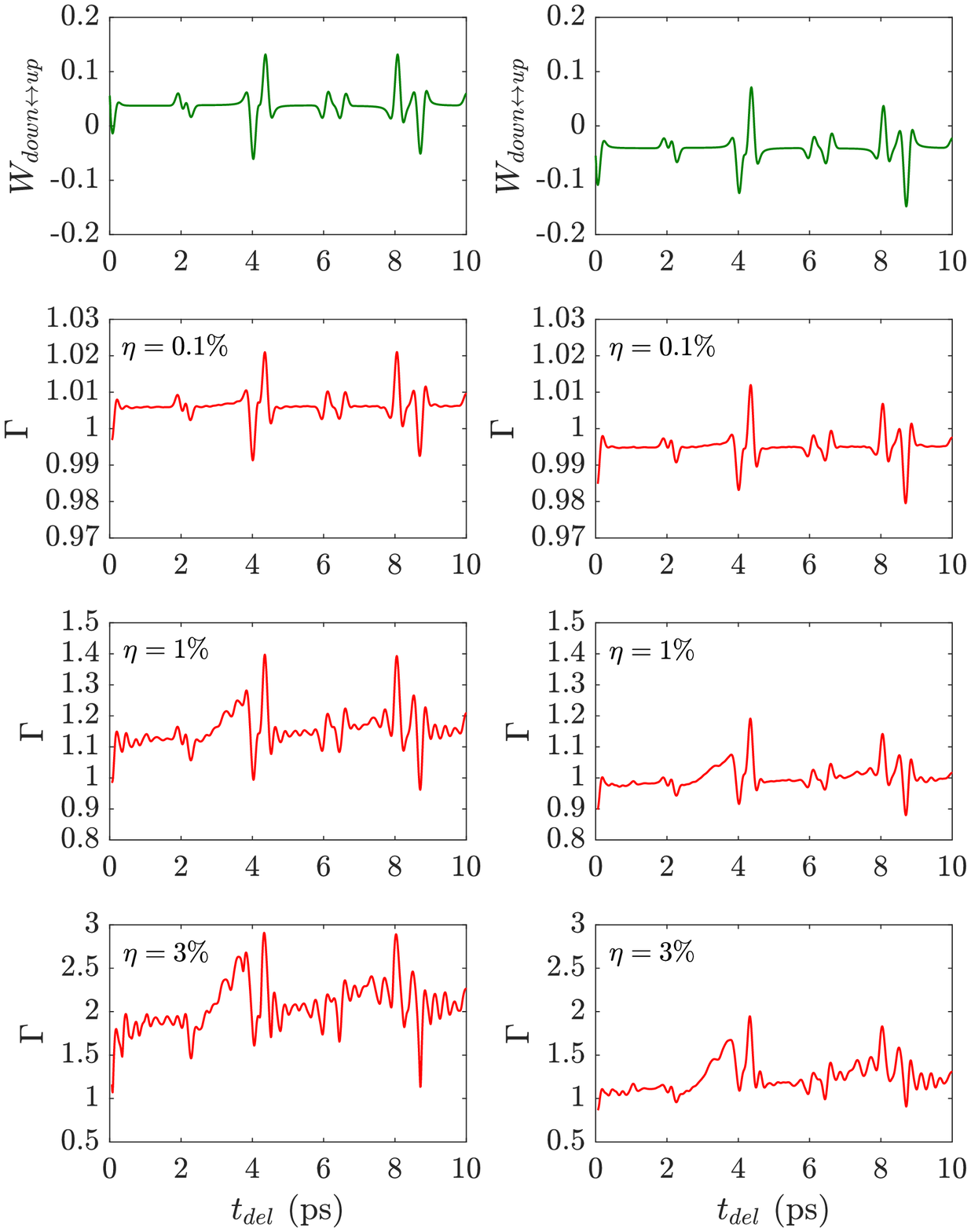}
	\caption{Gain/absorption defined according to the perturbative estimate
	Eq.~\eqref{Wtran}, and using the numerical spectrum to define the gain
	through Eq.~\eqref{Gain}. Left side
	corresponds to $p_X=0.45$ and $p_B=0.55$ (with electronic inversion), and the right side is for $p_X=0.55$
	and $p_B=0.45$ (no electronic inversion) . The results for $\Gamma$ are presented for
	various ionization levels $\eta=0.1\%$, 1$\%$, and 3$\%$ as noted. Other simulation
	parameters are $z_{max}=0.5$ mm, $\lambda_s=391$ nm, $\sigma_s = 20$ fs, $I_{pump}=10^{14}$ W/cm$^2$.}
	\label{FigGainDelay}
\end{figure}

Figure \ref{FigGainDelay} shows various cases of the total gain and absorption
as a function of the seed delay $t_{del}$.  The left column corresponds to
$p_X=0.45$ and $p_B=0.55$ where electronic inversion is present, while the
right column corresponds to $p_X=0.55$ and $p_B=0.45$ where electronic
inversion is absent. All other parameters are the same as used in
Fig.\ref{FigSeedBeforeAfter}.  The top row of Figure \ref{FigGainDelay} plots
the perturbative estimate $W_{down \leftrightarrow up}$, while the
following rows plot the $\Gamma$ parameter computed from the full numerical
propagation of the seed, for various values of the ionization fraction $\eta$
which are labeled on the plots.

The $W_{down \leftrightarrow up}$ curves shown in Figure
\ref{FigGainDelay} show that the expected gain and absorption is modulated as
the rotational wave packets on $X$ and $B$ evolve and modulate the $\langle
\cos^2\theta \rangle^{X,B}(t_{del})$ parameters that enter into the $W_{down
\leftrightarrow up}$ estimate.  Importantly, one can see that in both
cases of inversion (left) or no inversion (right) the behavior of the emission
can switch from gain to absorption and back depending on the particular delay
chosen to launch the seed pulse.  Regarding the results for the full seed
propagation, we can see that in the case of $\eta = 0.1\%$ (low ionization) the
numerically-calculated gain $\Gamma$ almost exactly mirrors the predictions of
the $W_{down \leftrightarrow up}$ estimate; when $W_{down \leftrightarrow up} >
0$ gain is predicted and correspondingly the results of the numerical
propagation of the seed yield $\Gamma>1$.  We emphasis that these results
demonstrate that gain can be achieved in the absence of electronic inversion
when there are rotational coherences present that can modulate the balance
between emission and absorption in the system.

\begin{figure}[t]
	\includegraphics[width=0.95\columnwidth]{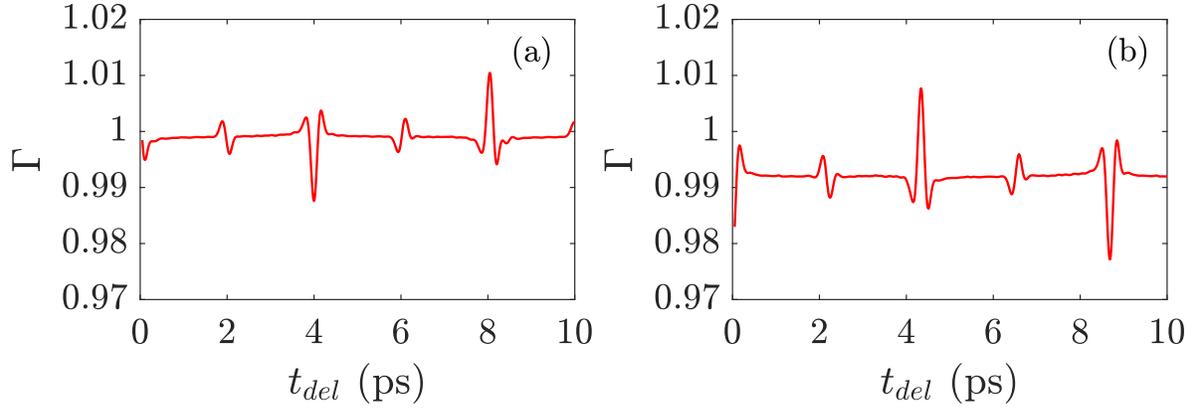}
	\caption{Gain/absorption as a function of pump-seed delay $t_{del}$
	for hypothetical cases where (a) the $X$ state was kept in a thermal 
	distribution of rotational states, and (b) for the case where
	$B$ was kept in a thermal distribution of rotational states.
	All other simulation parameters are the same as used 
	in Fig.\ref{FigGainDelay} with $p_X=0.55$ and $p_B=0.45$ (no electronic inversion).}
	\label{FigHypothetical}
\end{figure}

\begin{figure}[b]
	\includegraphics[width=0.95\columnwidth]{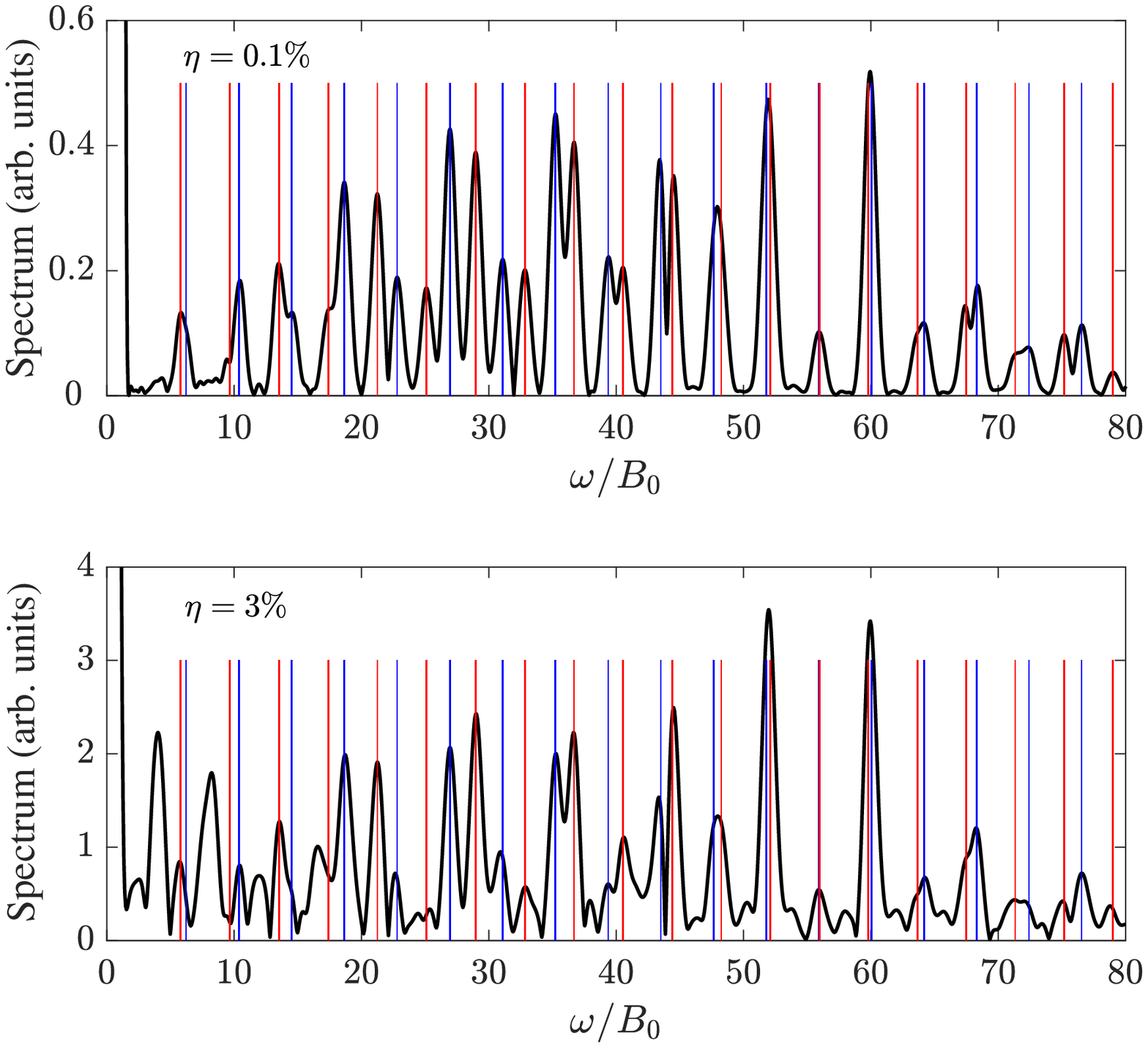}
	\caption{Fourier transform of the delay-dependent gain/absorption signal $\Gamma(t_{del})$
	for ionization fraction of $\eta=0.1\%$ and 3$\%$.  The frequency axis 
	has been normalized by the rotational constant of the neutral, which is
	labeled as $B_0$ on this plot.
	The red lines denote the expected positions of the $X$ state peaks
	$\omega^X_J = {\cal E}^X_{J+2} - {\cal E}^X_J$, while the blue lines denote the
	expected positions for the $B$ state $\omega^B_J = {\cal E}^B_{J+2} - {\cal E}^B_J$.  These simulations
	are for the same parameters used in Fig.\ref{FigGainDelay}, and
	correspond to the $p_X=0.45$ and $p_B=0.55$ case. }
	\label{FigFourierOfGain}
\end{figure}

In the cases of $\eta$ = 1\% and 3\% also shown in Fig.\ref{FigGainDelay}, we
see that $\Gamma$ starts to diverge from the $W_{down \leftrightarrow up}$
estimate.  This occurs because in these cases the density of the ions is large
enough to generate substantial gain in the seed, and as the amplitudes of the
gain lines in the seed spectrum increase the assumption of a flat spectrum
required to derive $W_{down \leftrightarrow up}$ no longer holds.  The
increased strength of the gain, and hence the increased amplitude of the
corresponding gain lines, is evidenced by the fact that $\Gamma$ reaches a
maximum of about 1.02 in the $\eta$ = 0.1\% case while it shoots up to about
1.38 and 2.9 in the $\eta$ = 1\% and 3\% cases respectively. Recall that due to
the definition of $\Gamma$ given in Eq.~(\ref{Gain}), a value close to 1
implies a small change in total intensity of the seed pulse, while a larger
value like 2.9 implies a rise in intensity of that same amount at the
$B\leftrightarrow X$ transition frequencies.
Deviations of $\Gamma$ away from $W_{down \leftrightarrow up}$ would
equivalently occur in the case of $\eta$ = 0.1\% if the propagation length is
increased; as the propagation length increases so will the amplitudes of the
gain lines, which in turn will cause a breakdown of the flat spectrum
approximation. Qualitatively, the deviations away from $W_{down \leftrightarrow
up}$ in the large gain regime appear as an increased amount
of oscillations in $\Gamma$ compared to what one would expect from the behaviors of 
$\langle \cos^2\theta\rangle^X(t)$ and $\langle \cos^2\theta\rangle^B(t)$ alone.
These increased oscillations in the delay-dependent gain
of a seed pulse have been observed in recent experiments \cite{Britton19}.

Figure \ref{FigHypothetical} shows additional results of the modulation of the
gain for the hypothetical cases where only one of the two ionic states was
rotationally pumped.  In Fig.\ref{FigHypothetical}a, we only generated a
rotational wave packet in the $B$ state while forcing the population in $X$
to be in a thermal distribution of rotational states at $T$=298 K.  The
relative ionic populations where set to $p_X=0.55$ and $p_B=0.45$ (i.e.
no electronic inversion) and $\eta$ = 0.1\% was used.  There remain values of
the delay where gain ($\Gamma>1$) is achieved.
Fig.\ref{FigHypothetical}b, shows the analogous case where we have kept the
coherent rotational excitations in $X$ while replacing the $B$ state with a
thermal distribution.  Again we can see delays where gain occurs.  These
simulations show that it is enough to have rotational coherences in only one of
the participating electronic states in order to generate gain without
inversion.

\begin{figure}[t]
	\includegraphics[width=0.95\columnwidth]{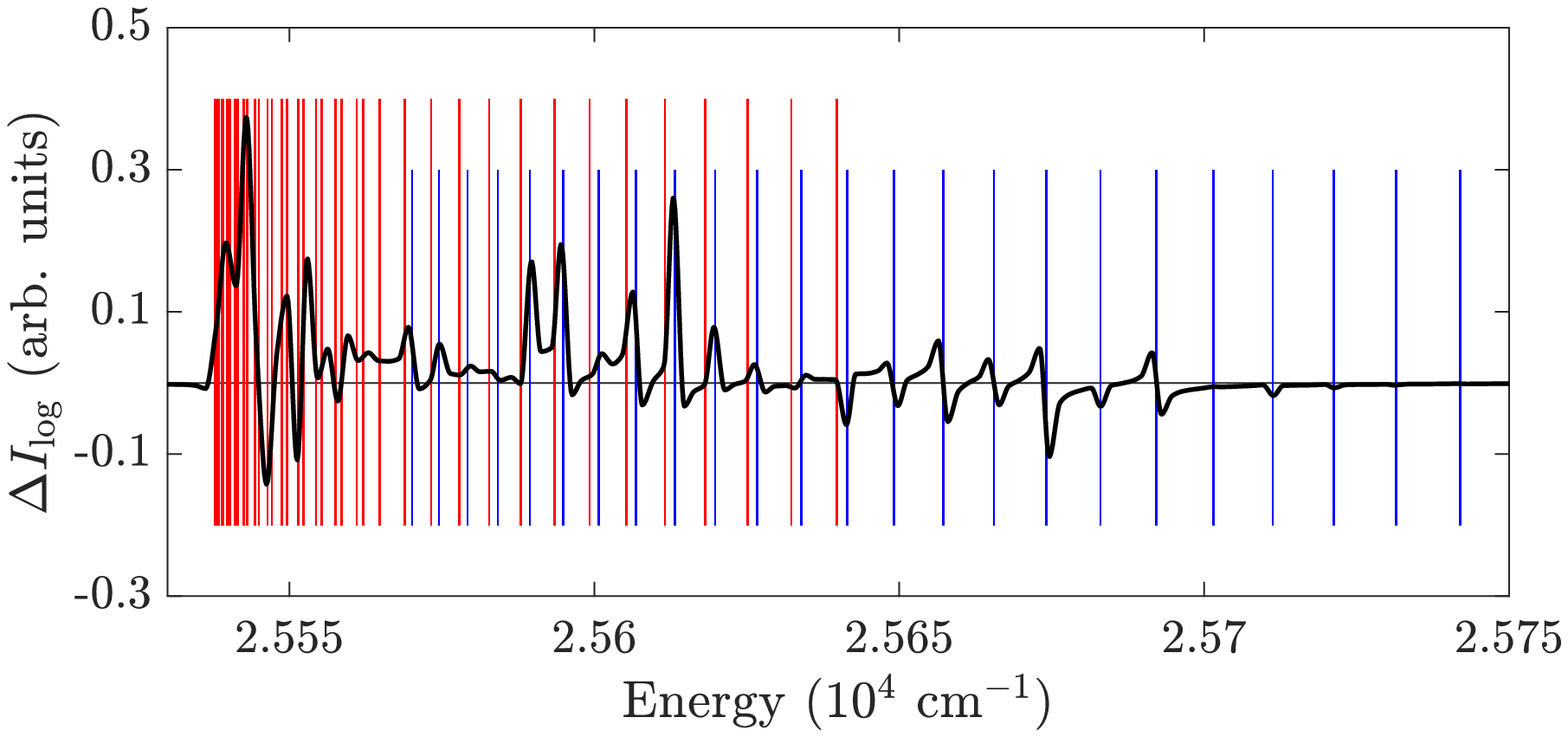}
	\caption{Gain and absorption lines in the output spectrum of the seed pulse 
	for a delay of $t_{del}$ = 4.3 ps.  All simulation parameters are
	the same as used in Fig.\ref{FigGainDelay} with $\eta=0.1$\%, $p_X=0.45$ and $p_B=0.55$. 
	Plotted is the difference between the output and input spectrum,
	$\Delta I_{\log}$ [see Eq.~(\ref{EqIlog})], across the range of energies
	corresponding to the $B\leftrightarrow X$ transitions.
	The red lines denote the P-branch transitions, while the blue lines
	show the R-branch transitions (see Fig.\ref{branches}).
	}\label{FigSeedZoom}
\end{figure}

Finally, we consider the Fourier spectrum of the delay-dependent gain signal.
It can be shown that, for linear molecules, the alignment parameter $\langle
\cos^2\theta \rangle(t)$ contains the frequencies 
\begin{equation}\label{Eq4Jp6}
	\omega_J = {\cal E}_{J+2} - {\cal E}_J \approx B(4J+6),
\end{equation}
where $B$ is the rotational constant of the molecule being considered
\cite{Dooley03}.  In the low gain regime, we have seen that the
delay-dependent emission/absorption temporally follows the formula in Eq.~(\ref{Wtran})
and hence we expect the Fourier transform of the gain/absorption to contain two
series of peaks like in Eq.~(\ref{Eq4Jp6}), one reflecting the rotational
spacings of $X$ and the other reflecting the rotational spacings of $B$.
Fig.\ref{FigFourierOfGain} shows the Fourier transform for two of the
absorption/gain signals $\Gamma(t_{del})$ presented in Fig.~\ref{FigGainDelay},
one in the low-gain regime ($\eta=0.1\%$) and the second for the high-gain
regime ($\eta=3\%$).
In the low-gain regime ($\eta=0.1\%$), two
series of peaks in the Fourier spectrum can be seen, and they line up perfectly
with the expected frequencies $\omega^k_J = {\cal E}^k_{J+2} - {\cal E}^k_J$ ($k={X,B}$) for 
the $X$ and $B$ states.  However, once the gain becomes larger and
the delay-dependent emission diverges from the $W_{down \leftrightarrow
up}$ estimate, new frequencies that are not accounted for by these $\omega^k_J$
arise.  This can be seen in the $\eta=3\%$
case in Fig.\ref{FigFourierOfGain};
new frequencies that do not align with $\omega^k_J$
are now present.  These new frequencies are a result of
the interplay between the timescales required for the gain lines to grow
substantially in amplitude and the timescales of the coherent rotational
wave packet.

%%%%%%%%%%%%%%%%%%%%%%%%%%%%%%%%%%%%%%%%%%%%%%%%%%%%%%%%%%%%%%%%%%%%%%%%%%%%
\subsection{Structure of the spectra}

\begin{figure}[t]
	\includegraphics[width=0.95\columnwidth]{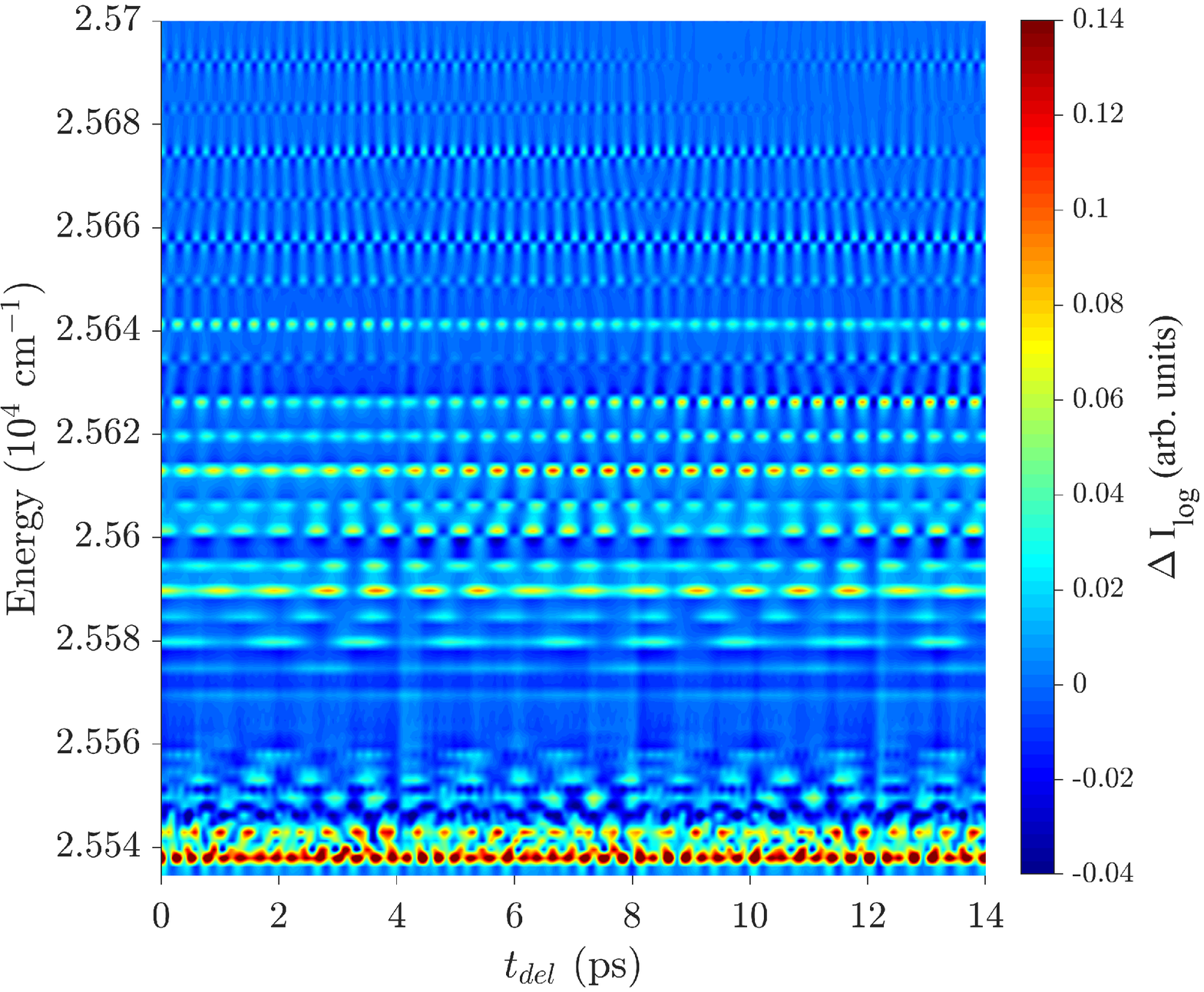}
	\caption{Delay dependence of the gain and absorption lines in the output seed spectrum.
	As in Fig.~\ref{FigSeedZoom},  $\Delta I_{\log}$ [see Eq.~(\ref{EqIlog})]
	is plotted across the $B\leftrightarrow X$ transition region.}
	\label{FigLadan2D}
\end{figure}

Figures \ref{FigSeedZoom} and \ref{FigLadan2D} present a more detailed view of
the gain and absorption lines in the output seed pulse.  Both figures plot 
a scaled logarithm of the change in the output and input spectrum of the seed defined by
\begin{equation}\label{EqIlog}
	\Delta  I_{\log} = \log_{10} \left( \frac{ \Delta I(\omega) + \Delta I_{min} + \Delta I_{max}}{ \Delta I_{min} + \Delta I_{max} } \right)
\end{equation}
where $\Delta I(\omega) = I_{out}(\omega) - I_{in}(\omega)$ is the difference
between the output and input spectral intensities of the seed, and $\Delta
I_{min}$ and $\Delta I_{max}$ are the minimum and maximum of $\Delta I(\omega)$
respectively.  The range of energies plotted  corresponds to the energy window
of the $B\leftrightarrow X$ transitions.  With the definition in
Eq~(\ref{EqIlog}), $\Delta I_{\log} > 0$ represents gain, while $\Delta
I_{\log} < 0$ corresponds to absorption.  All simulation parameters are the
same as used in Fig.\ref{FigGainDelay} with $\eta = 0.1$\%, $p_X=0.45$ and $p_B=0.55$.
Fig.\ref{FigSeedZoom} shows $\Delta I_{\log}$ for the specific delay of
$t_{del}$ = 4.3 ps, while Fig.~\ref{FigLadan2D} shows $\Delta I_{\log}$ for a
range of delays;  Fig.~\ref{FigSeedZoom} is a slice through
Fig.~\ref{FigLadan2D} at the delay $t_{del}$ = 4.3 ps.  The 2D spectrum in
Fig.~\ref{FigLadan2D} exhibits a rich modulation structure that is a result of
the underlying rotational coherences comprising the rotational wave packets.
These structures and modulations are in excellent agreement with those found
experimentally in high-resolution measurements of the delay-dependent seeded
N$_2^+$ lasing as can be seen by comparing against Fig.1 from
Ref.\cite{Arissian18}, which confirms that our model is capturing the essential
effects of the rotational coherences on the gain and absorption of the delayed
seed pulse.  One final comment regarding the spectrum in Fig.~\ref{FigLadan2D}
relates to the subtle vertical features that appear near 4, 8, and 12 ps.
These features are caused by the time-dependent refractive index of the neutral
rotational wavepackets [Eqs.~(\ref{refract}) and (\ref{theta_pump})].  Similar
modulations appearing at the revival times of the neutral rotational
wavepackets have also been observed, and these modulations caused by the
time-dependent refractive index of the neutral where experimentally found to
persist long after the ion-driven rotational modulations of the gain have
decayed away \cite{Arissian18}.

%%%%%%%%%%%%%%%%%%%%%%%%%%%%%%%%%%%%%%%%%%%%%%%%%%%%%%%%%%%%%%%%%%%%%%%%%%%%%%%%%
%%%%%%%%%%%%%%%%%%%%%%%%%%%%%%%%%%%%%%%%%%%%%%%%%%%%%%%%%%%%%%%%%%%%%%%%%%%%%%%%%
\section{Conclusion} 
Motivated by the seeded version of ultrafast N$_2^+$ lasing, we have developed
and explored a coupled Maxwell/von Neumann model that simulates the
propagation of a seed pulse in an ionized and rotationally-excited gas of N$_2$
molecules.  The model is aimed at understanding how the presence of rotational
wavepackets modulates the absorption and gain of a delayed seed pulse and
expands on the idea of transient inversion in rotationally-aligned nitrogen
first discussed in Ref.\cite{Kartashov14}.  Our numerical results successfully
capture experimentally-observed modulations \cite{Zhang13,Arissian18,Britton19}
of delay-dependent gain properties driven by the coherent rotational
excitations.  Using first-order time-dependent perturbation theory, we show
that the gain estimate $W_{down \leftrightarrow up}$ suggested in
Ref.\cite{Kartashov14} correctly captures the rotational modulations of the
gain/absorption in the limit of low gain and absorption.  As the total gain
increases, the numerically-calculated rotationally-driven modulations of the
gain start to diverge from the $W_{down \leftrightarrow up}$ estimate. This
divergent behavior occurs due to the non-flat structure of the seed spectrum
that results from the growth of the gain lines at the $B\leftrightarrow X$
transition frequencies.  Finally, we have demonstrated that gain in the absence
of electronic inversion is possible in N$_2^+$ lasing due to the presence of
rotational coherence.

\section{Acknowledgments}
We thank Paul Corkum, Ladan Arissian, Mathew Britton, and David Villeneuve for
numerous stimulating discussion regarding N$_2^+$ lasing.  M.S. acknowledges
the financial support from the Natural Science and Engineering Research Council
(NSERC) of Canada through their Discovery Grants program.

%%%%%%%%%%%%%%%%%%%%%%%%%%%%%%%%%%%%%%%%%%%%%%%%%%%%%%%%%%%%%%%%%%%%%%%%%%%%%%%%%
%%%%%%%%%%%%%%%%%%%%%%%%%%%%%%%%%%%%%%%%%%%%%%%%%%%%%%%%%%%%%%%%%%%%%%%%%%%%%%%%%
%\bibliography{biblio}

\begin{thebibliography}{40}%
\makeatletter
\providecommand \@ifxundefined [1]{%
 \@ifx{#1\undefined}
}%
\providecommand \@ifnum [1]{%
 \ifnum #1\expandafter \@firstoftwo
 \else \expandafter \@secondoftwo
 \fi
}%
\providecommand \@ifx [1]{%
 \ifx #1\expandafter \@firstoftwo
 \else \expandafter \@secondoftwo
 \fi
}%
\providecommand \natexlab [1]{#1}%
\providecommand \enquote  [1]{``#1''}%
\providecommand \bibnamefont  [1]{#1}%
\providecommand \bibfnamefont [1]{#1}%
\providecommand \citenamefont [1]{#1}%
\providecommand \href@noop [0]{\@secondoftwo}%
\providecommand \href [0]{\begingroup \@sanitize@url \@href}%
\providecommand \@href[1]{\@@startlink{#1}\@@href}%
\providecommand \@@href[1]{\endgroup#1\@@endlink}%
\providecommand \@sanitize@url [0]{\catcode `\\12\catcode `\$12\catcode
  `\&12\catcode `\#12\catcode `\^12\catcode `\_12\catcode `\%12\relax}%
\providecommand \@@startlink[1]{}%
\providecommand \@@endlink[0]{}%
\providecommand \url  [0]{\begingroup\@sanitize@url \@url }%
\providecommand \@url [1]{\endgroup\@href {#1}{\urlprefix }}%
\providecommand \urlprefix  [0]{URL }%
\providecommand \Eprint [0]{\href }%
\providecommand \doibase [0]{http://dx.doi.org/}%
\providecommand \selectlanguage [0]{\@gobble}%
\providecommand \bibinfo  [0]{\@secondoftwo}%
\providecommand \bibfield  [0]{\@secondoftwo}%
\providecommand \translation [1]{[#1]}%
\providecommand \BibitemOpen [0]{}%
\providecommand \bibitemStop [0]{}%
\providecommand \bibitemNoStop [0]{.\EOS\space}%
\providecommand \EOS [0]{\spacefactor3000\relax}%
\providecommand \BibitemShut  [1]{\csname bibitem#1\endcsname}%
\let\auto@bib@innerbib\@empty
%</preamble>
\bibitem [{\citenamefont {Friedrich}\ and\ \citenamefont
  {Herschbach}(1995)}]{Friedrich95}%
  \BibitemOpen
  \bibfield  {author} {\bibinfo {author} {\bibfnamefont {B.}~\bibnamefont
  {Friedrich}}\ and\ \bibinfo {author} {\bibfnamefont {D.}~\bibnamefont
  {Herschbach}},\ }\href {www.scopus.com} {\bibfield  {journal} {\bibinfo
  {journal} {Phys. Rev. Lett.}\ }\textbf {\bibinfo {volume} {74}},\ \bibinfo
  {pages} {4623} (\bibinfo {year} {1995})}\BibitemShut {NoStop}%
\bibitem [{\citenamefont {Larsen}\ \emph {et~al.}(1999)\citenamefont {Larsen},
  \citenamefont {Wendt-Larsen},\ and\ \citenamefont {Stapelfeldt}}]{Larsen99}%
  \BibitemOpen
  \bibfield  {author} {\bibinfo {author} {\bibfnamefont {J.}~\bibnamefont
  {Larsen}}, \bibinfo {author} {\bibfnamefont {I.}~\bibnamefont
  {Wendt-Larsen}}, \ and\ \bibinfo {author} {\bibfnamefont {H.}~\bibnamefont
  {Stapelfeldt}},\ }\href@noop {} {\bibfield  {journal} {\bibinfo  {journal}
  {Phys. Rev. Lett.}\ }\textbf {\bibinfo {volume} {83}},\ \bibinfo {pages}
  {1123} (\bibinfo {year} {1999})}\BibitemShut {NoStop}%
\bibitem [{\citenamefont {Stapelfeldt}\ and\ \citenamefont
  {Seideman}(2003)}]{Stapel03}%
  \BibitemOpen
  \bibfield  {author} {\bibinfo {author} {\bibfnamefont {H.}~\bibnamefont
  {Stapelfeldt}}\ and\ \bibinfo {author} {\bibfnamefont {T.}~\bibnamefont
  {Seideman}},\ }\href {www.scopus.com} {\bibfield  {journal} {\bibinfo
  {journal} {Rev. Mod. Phys.}\ }\textbf {\bibinfo {volume} {75}},\ \bibinfo
  {pages} {543} (\bibinfo {year} {2003})}\BibitemShut {NoStop}%
\bibitem [{\citenamefont {Dooley}\ \emph {et~al.}(2003)\citenamefont {Dooley},
  \citenamefont {Litvinyuk}, \citenamefont {Lee}, \citenamefont {Rayner},
  \citenamefont {Spanner}, \citenamefont {Villeneuve},\ and\ \citenamefont
  {Corkum}}]{Dooley03}%
  \BibitemOpen
  \bibfield  {author} {\bibinfo {author} {\bibfnamefont {P.~W.}\ \bibnamefont
  {Dooley}}, \bibinfo {author} {\bibfnamefont {I.~V.}\ \bibnamefont
  {Litvinyuk}}, \bibinfo {author} {\bibfnamefont {K.~F.}\ \bibnamefont {Lee}},
  \bibinfo {author} {\bibfnamefont {D.~M.}\ \bibnamefont {Rayner}}, \bibinfo
  {author} {\bibfnamefont {M.}~\bibnamefont {Spanner}}, \bibinfo {author}
  {\bibfnamefont {D.~M.}\ \bibnamefont {Villeneuve}}, \ and\ \bibinfo {author}
  {\bibfnamefont {P.~B.}\ \bibnamefont {Corkum}},\ }\href@noop {} {\bibfield
  {journal} {\bibinfo  {journal} {Phys. Rev. A}\ }\textbf {\bibinfo {volume}
  {68}},\ \bibinfo {pages} {234061} (\bibinfo {year} {2003})}\BibitemShut
  {NoStop}%
\bibitem [{\citenamefont {Seideman}(1999)}]{Seideman99}%
  \BibitemOpen
  \bibfield  {author} {\bibinfo {author} {\bibfnamefont {T.}~\bibnamefont
  {Seideman}},\ }\href@noop {} {\bibfield  {journal} {\bibinfo  {journal}
  {Phys. Rev. Lett.}\ }\textbf {\bibinfo {volume} {83}},\ \bibinfo {pages}
  {4971} (\bibinfo {year} {1999})}\BibitemShut {NoStop}%
\bibitem [{\citenamefont {Ortigoso}\ \emph {et~al.}(1999)\citenamefont
  {Ortigoso}, \citenamefont {Rodriguez}, \citenamefont {Gupta},\ and\
  \citenamefont {Friedrich}}]{Ortigoso99}%
  \BibitemOpen
  \bibfield  {author} {\bibinfo {author} {\bibfnamefont {J.}~\bibnamefont
  {Ortigoso}}, \bibinfo {author} {\bibfnamefont {M.}~\bibnamefont {Rodriguez}},
  \bibinfo {author} {\bibfnamefont {M.}~\bibnamefont {Gupta}}, \ and\ \bibinfo
  {author} {\bibfnamefont {B.}~\bibnamefont {Friedrich}},\ }\href@noop {}
  {\bibfield  {journal} {\bibinfo  {journal} {J. Chem. Phys.}\ }\textbf
  {\bibinfo {volume} {110}},\ \bibinfo {pages} {3870} (\bibinfo {year}
  {1999})}\BibitemShut {NoStop}%
\bibitem [{\citenamefont {Rosca-Pruna}\ and\ \citenamefont
  {Vrakking}(2001)}]{Rosca01}%
  \BibitemOpen
  \bibfield  {author} {\bibinfo {author} {\bibfnamefont {F.}~\bibnamefont
  {Rosca-Pruna}}\ and\ \bibinfo {author} {\bibfnamefont {M.}~\bibnamefont
  {Vrakking}},\ }\href@noop {} {\bibfield  {journal} {\bibinfo  {journal}
  {Phys. Rev. Lett.}\ }\textbf {\bibinfo {volume} {87}},\ \bibinfo {pages}
  {153902} (\bibinfo {year} {2001})}\BibitemShut {NoStop}%
\bibitem [{\citenamefont {Eberly}\ \emph {et~al.}(1980)\citenamefont {Eberly},
  \citenamefont {Narozhny},\ and\ \citenamefont {Sanchez-Mondragon}}]{Eberly}%
  \BibitemOpen
  \bibfield  {author} {\bibinfo {author} {\bibfnamefont {J.}~\bibnamefont
  {Eberly}}, \bibinfo {author} {\bibfnamefont {N.}~\bibnamefont {Narozhny}}, \
  and\ \bibinfo {author} {\bibfnamefont {J.}~\bibnamefont
  {Sanchez-Mondragon}},\ }\href@noop {} {\bibfield  {journal} {\bibinfo
  {journal} {Phys. Rev. Lett.}\ }\textbf {\bibinfo {volume} {44}},\ \bibinfo
  {pages} {1323} (\bibinfo {year} {1980})}\BibitemShut {NoStop}%
\bibitem [{\citenamefont {Averbukh}\ and\ \citenamefont
  {Perelman}(1989)}]{Averb}%
  \BibitemOpen
  \bibfield  {author} {\bibinfo {author} {\bibfnamefont {I.}~\bibnamefont
  {Averbukh}}\ and\ \bibinfo {author} {\bibfnamefont {N.}~\bibnamefont
  {Perelman}},\ }\href@noop {} {\bibfield  {journal} {\bibinfo  {journal}
  {Phys. Lett. A}\ }\textbf {\bibinfo {volume} {139}},\ \bibinfo {pages} {449}
  (\bibinfo {year} {1989})}\BibitemShut {NoStop}%
\bibitem [{\citenamefont {Kalosha}\ \emph {et~al.}(2002)\citenamefont
  {Kalosha}, \citenamefont {Spanner}, \citenamefont {Herrmann},\ and\
  \citenamefont {Ivanov}}]{Kalosha02}%
  \BibitemOpen
  \bibfield  {author} {\bibinfo {author} {\bibfnamefont {V.}~\bibnamefont
  {Kalosha}}, \bibinfo {author} {\bibfnamefont {M.}~\bibnamefont {Spanner}},
  \bibinfo {author} {\bibfnamefont {J.}~\bibnamefont {Herrmann}}, \ and\
  \bibinfo {author} {\bibfnamefont {M.}~\bibnamefont {Ivanov}},\ }\href@noop {}
  {\bibfield  {journal} {\bibinfo  {journal} {Phys. Rev. Lett.}\ }\textbf
  {\bibinfo {volume} {88}},\ \bibinfo {pages} {103901} (\bibinfo {year}
  {2002})}\BibitemShut {NoStop}%
\bibitem [{\citenamefont {Bartels}\ \emph {et~al.}(2002)\citenamefont
  {Bartels}, \citenamefont {Weinacht}, \citenamefont {Wagner}, \citenamefont
  {Baertschy}, \citenamefont {Greene}, \citenamefont {Murnane},\ and\
  \citenamefont {Kapteyn}}]{Bartels02}%
  \BibitemOpen
  \bibfield  {author} {\bibinfo {author} {\bibfnamefont {R.}~\bibnamefont
  {Bartels}}, \bibinfo {author} {\bibfnamefont {T.}~\bibnamefont {Weinacht}},
  \bibinfo {author} {\bibfnamefont {N.}~\bibnamefont {Wagner}}, \bibinfo
  {author} {\bibfnamefont {M.}~\bibnamefont {Baertschy}}, \bibinfo {author}
  {\bibfnamefont {C.}~\bibnamefont {Greene}}, \bibinfo {author} {\bibfnamefont
  {M.}~\bibnamefont {Murnane}}, \ and\ \bibinfo {author} {\bibfnamefont
  {H.}~\bibnamefont {Kapteyn}},\ }\href@noop {} {\bibfield  {journal} {\bibinfo
   {journal} {Phys. Rev. Lett.}\ }\textbf {\bibinfo {volume} {88}},\ \bibinfo
  {pages} {2002} (\bibinfo {year} {2002})}\BibitemShut {NoStop}%
\bibitem [{\citenamefont {Thekkadath}\ \emph {et~al.}(2016)\citenamefont
  {Thekkadath}, \citenamefont {Heshami}, \citenamefont {England}, \citenamefont
  {Bustard}, \citenamefont {Sussman},\ and\ \citenamefont {Spanner}}]{Thekk16}%
  \BibitemOpen
  \bibfield  {author} {\bibinfo {author} {\bibfnamefont {G.~S.}\ \bibnamefont
  {Thekkadath}}, \bibinfo {author} {\bibfnamefont {K.}~\bibnamefont {Heshami}},
  \bibinfo {author} {\bibfnamefont {D.~G.}\ \bibnamefont {England}}, \bibinfo
  {author} {\bibfnamefont {P.~J.}\ \bibnamefont {Bustard}}, \bibinfo {author}
  {\bibfnamefont {B.~J.}\ \bibnamefont {Sussman}}, \ and\ \bibinfo {author}
  {\bibfnamefont {M.}~\bibnamefont {Spanner}},\ }\href {www.scopus.com}
  {\bibfield  {journal} {\bibinfo  {journal} {J. Mod. Opt.}\ }\textbf {\bibinfo
  {volume} {63}},\ \bibinfo {pages} {2093} (\bibinfo {year}
  {2016})}\BibitemShut {NoStop}%
\bibitem [{\citenamefont {Luo}\ \emph {et~al.}(2003)\citenamefont {Luo},
  \citenamefont {Liu},\ and\ \citenamefont {Chin}}]{Chin03}%
  \BibitemOpen
  \bibfield  {author} {\bibinfo {author} {\bibfnamefont {Q.}~\bibnamefont
  {Luo}}, \bibinfo {author} {\bibfnamefont {W.}~\bibnamefont {Liu}}, \ and\
  \bibinfo {author} {\bibfnamefont {S.}~\bibnamefont {Chin}},\ }\href@noop {}
  {\bibfield  {journal} {\bibinfo  {journal} {App. Phys. B}\ }\textbf {\bibinfo
  {volume} {76}},\ \bibinfo {pages} {337} (\bibinfo {year} {2003})}\BibitemShut
  {NoStop}%
\bibitem [{\citenamefont {Yao}\ \emph {et~al.}(2011)\citenamefont {Yao},
  \citenamefont {Zeng}, \citenamefont {Xu}, \citenamefont {Li}, \citenamefont
  {Chu}, \citenamefont {Ni}, \citenamefont {Zhang}, \citenamefont {Chin},
  \citenamefont {Cheng},\ and\ \citenamefont {Xu}}]{Yao11}%
  \BibitemOpen
  \bibfield  {author} {\bibinfo {author} {\bibfnamefont {J.}~\bibnamefont
  {Yao}}, \bibinfo {author} {\bibfnamefont {B.}~\bibnamefont {Zeng}}, \bibinfo
  {author} {\bibfnamefont {H.}~\bibnamefont {Xu}}, \bibinfo {author}
  {\bibfnamefont {G.}~\bibnamefont {Li}}, \bibinfo {author} {\bibfnamefont
  {W.}~\bibnamefont {Chu}}, \bibinfo {author} {\bibfnamefont {J.}~\bibnamefont
  {Ni}}, \bibinfo {author} {\bibfnamefont {H.}~\bibnamefont {Zhang}}, \bibinfo
  {author} {\bibfnamefont {S.}~\bibnamefont {Chin}}, \bibinfo {author}
  {\bibfnamefont {Y.}~\bibnamefont {Cheng}}, \ and\ \bibinfo {author}
  {\bibfnamefont {Z.}~\bibnamefont {Xu}},\ }\href@noop {} {\bibfield  {journal}
  {\bibinfo  {journal} {Rhys. Rev. A}\ }\textbf {\bibinfo {volume} {84}},\
  \bibinfo {pages} {051802(R)} (\bibinfo {year} {2011})}\BibitemShut {NoStop}%
\bibitem [{\citenamefont {Liu}\ \emph {et~al.}(2013)\citenamefont {Liu},
  \citenamefont {Brelet}, \citenamefont {Point}, \citenamefont {Houard},\ and\
  \citenamefont {Mysyrowicz}}]{Liu13}%
  \BibitemOpen
  \bibfield  {author} {\bibinfo {author} {\bibfnamefont {Y.}~\bibnamefont
  {Liu}}, \bibinfo {author} {\bibfnamefont {Y.}~\bibnamefont {Brelet}},
  \bibinfo {author} {\bibfnamefont {G.}~\bibnamefont {Point}}, \bibinfo
  {author} {\bibfnamefont {A.}~\bibnamefont {Houard}}, \ and\ \bibinfo {author}
  {\bibnamefont {Mysyrowicz}},\ }\href@noop {} {\bibfield  {journal} {\bibinfo
  {journal} {Opt. Exp.}\ }\textbf {\bibinfo {volume} {21}},\ \bibinfo {pages}
  {22792} (\bibinfo {year} {2013})}\BibitemShut {NoStop}%
\bibitem [{\citenamefont {Ni}\ \emph {et~al.}(2013)\citenamefont {Ni},
  \citenamefont {Chu}, \citenamefont {Jing}, \citenamefont {Zhang},
  \citenamefont {Zeng}, \citenamefont {Yao}, \citenamefont {Li}, \citenamefont
  {Xie}, \citenamefont {Zhang}, \citenamefont {Xu}, \citenamefont {Chin},
  \citenamefont {Cheng},\ and\ \citenamefont {Xu}}]{Ni13}%
  \BibitemOpen
  \bibfield  {author} {\bibinfo {author} {\bibfnamefont {J.}~\bibnamefont
  {Ni}}, \bibinfo {author} {\bibfnamefont {W.}~\bibnamefont {Chu}}, \bibinfo
  {author} {\bibfnamefont {C.}~\bibnamefont {Jing}}, \bibinfo {author}
  {\bibfnamefont {H.}~\bibnamefont {Zhang}}, \bibinfo {author} {\bibfnamefont
  {B.}~\bibnamefont {Zeng}}, \bibinfo {author} {\bibfnamefont {J.}~\bibnamefont
  {Yao}}, \bibinfo {author} {\bibfnamefont {G.}~\bibnamefont {Li}}, \bibinfo
  {author} {\bibfnamefont {H.}~\bibnamefont {Xie}}, \bibinfo {author}
  {\bibfnamefont {C.}~\bibnamefont {Zhang}}, \bibinfo {author} {\bibfnamefont
  {H.}~\bibnamefont {Xu}}, \bibinfo {author} {\bibfnamefont {S.}~\bibnamefont
  {Chin}}, \bibinfo {author} {\bibfnamefont {Y.}~\bibnamefont {Cheng}}, \ and\
  \bibinfo {author} {\bibfnamefont {Z.}~\bibnamefont {Xu}},\ }\href@noop {}
  {\bibfield  {journal} {\bibinfo  {journal} {Opt. Express}\ }\textbf {\bibinfo
  {volume} {21}},\ \bibinfo {pages} {8746} (\bibinfo {year}
  {2013})}\BibitemShut {NoStop}%
\bibitem [{\citenamefont {Zhang}\ \emph {et~al.}(2013)\citenamefont {Zhang},
  \citenamefont {Jing}, \citenamefont {Yao}, \citenamefont {Li}, \citenamefont
  {Zeng}, \citenamefont {Chu}, \citenamefont {Ni}, \citenamefont {Xie},
  \citenamefont {Xu}, \citenamefont {Chin}, \citenamefont {Yamanouchi},
  \citenamefont {Cheng},\ and\ \citenamefont {Xu}}]{Zhang13}%
  \BibitemOpen
  \bibfield  {author} {\bibinfo {author} {\bibfnamefont {H.}~\bibnamefont
  {Zhang}}, \bibinfo {author} {\bibfnamefont {C.}~\bibnamefont {Jing}},
  \bibinfo {author} {\bibfnamefont {J.}~\bibnamefont {Yao}}, \bibinfo {author}
  {\bibfnamefont {G.}~\bibnamefont {Li}}, \bibinfo {author} {\bibfnamefont
  {B.}~\bibnamefont {Zeng}}, \bibinfo {author} {\bibfnamefont {W.}~\bibnamefont
  {Chu}}, \bibinfo {author} {\bibfnamefont {J.}~\bibnamefont {Ni}}, \bibinfo
  {author} {\bibfnamefont {H.}~\bibnamefont {Xie}}, \bibinfo {author}
  {\bibfnamefont {H.}~\bibnamefont {Xu}}, \bibinfo {author} {\bibfnamefont
  {S.}~\bibnamefont {Chin}}, \bibinfo {author} {\bibfnamefont {K.}~\bibnamefont
  {Yamanouchi}}, \bibinfo {author} {\bibfnamefont {Y.}~\bibnamefont {Cheng}}, \
  and\ \bibinfo {author} {\bibfnamefont {Z.}~\bibnamefont {Xu}},\ }\href@noop
  {} {\bibfield  {journal} {\bibinfo  {journal} {Phys. Rev. X}\ }\textbf
  {\bibinfo {volume} {3}},\ \bibinfo {pages} {041009} (\bibinfo {year}
  {2013})}\BibitemShut {NoStop}%
\bibitem [{\citenamefont {Zeng}\ \emph {et~al.}(2014)\citenamefont {Zeng},
  \citenamefont {Chu}, \citenamefont {Li}, \citenamefont {Yao}, \citenamefont
  {Zhang}, \citenamefont {Ni}, \citenamefont {Jing}, \citenamefont {Xie},\ and\
  \citenamefont {Cheng}}]{Zeng14}%
  \BibitemOpen
  \bibfield  {author} {\bibinfo {author} {\bibfnamefont {B.}~\bibnamefont
  {Zeng}}, \bibinfo {author} {\bibfnamefont {W.}~\bibnamefont {Chu}}, \bibinfo
  {author} {\bibfnamefont {G.}~\bibnamefont {Li}}, \bibinfo {author}
  {\bibfnamefont {J.}~\bibnamefont {Yao}}, \bibinfo {author} {\bibfnamefont
  {H.}~\bibnamefont {Zhang}}, \bibinfo {author} {\bibfnamefont
  {J.}~\bibnamefont {Ni}}, \bibinfo {author} {\bibfnamefont {C.}~\bibnamefont
  {Jing}}, \bibinfo {author} {\bibfnamefont {H.}~\bibnamefont {Xie}}, \ and\
  \bibinfo {author} {\bibfnamefont {Y.}~\bibnamefont {Cheng}},\ }\href@noop {}
  {\bibfield  {journal} {\bibinfo  {journal} {Phys. Rev. A}\ }\textbf {\bibinfo
  {volume} {89}},\ \bibinfo {pages} {042508} (\bibinfo {year}
  {2014})}\BibitemShut {NoStop}%
\bibitem [{\citenamefont {Xu}\ \emph {et~al.}(2015)\citenamefont {Xu},
  \citenamefont {Ltstedt}, \citenamefont {Iwasaki},\ and\ \citenamefont
  {Yamanouchi}}]{Xu15}%
  \BibitemOpen
  \bibfield  {author} {\bibinfo {author} {\bibfnamefont {H.}~\bibnamefont
  {Xu}}, \bibinfo {author} {\bibfnamefont {E.}~\bibnamefont {Ltstedt}},
  \bibinfo {author} {\bibfnamefont {A.}~\bibnamefont {Iwasaki}}, \ and\
  \bibinfo {author} {\bibfnamefont {K.}~\bibnamefont {Yamanouchi}},\
  }\href@noop {} {\bibfield  {journal} {\bibinfo  {journal} {Nat. Commun.}\
  }\textbf {\bibinfo {volume} {6}},\ \bibinfo {pages} {8347} (\bibinfo {year}
  {2015})}\BibitemShut {NoStop}%
\bibitem [{\citenamefont {Yao}\ \emph {et~al.}(2016)\citenamefont {Yao},
  \citenamefont {Jiang}, \citenamefont {Chu}, \citenamefont {Zeng},
  \citenamefont {Wu}, \citenamefont {Lu}, \citenamefont {Li}, \citenamefont
  {Xie}, \citenamefont {Li}, \citenamefont {Yu}, \citenamefont {Wang},
  \citenamefont {Jiang}, \citenamefont {Gong},\ and\ \citenamefont
  {Cheng}}]{Yao16}%
  \BibitemOpen
  \bibfield  {author} {\bibinfo {author} {\bibfnamefont {J.}~\bibnamefont
  {Yao}}, \bibinfo {author} {\bibfnamefont {S.}~\bibnamefont {Jiang}}, \bibinfo
  {author} {\bibfnamefont {W.}~\bibnamefont {Chu}}, \bibinfo {author}
  {\bibfnamefont {B.}~\bibnamefont {Zeng}}, \bibinfo {author} {\bibfnamefont
  {C.}~\bibnamefont {Wu}}, \bibinfo {author} {\bibfnamefont {R.}~\bibnamefont
  {Lu}}, \bibinfo {author} {\bibfnamefont {Z.}~\bibnamefont {Li}}, \bibinfo
  {author} {\bibfnamefont {H.}~\bibnamefont {Xie}}, \bibinfo {author}
  {\bibfnamefont {G.}~\bibnamefont {Li}}, \bibinfo {author} {\bibfnamefont
  {C.}~\bibnamefont {Yu}}, \bibinfo {author} {\bibfnamefont {Z.}~\bibnamefont
  {Wang}}, \bibinfo {author} {\bibfnamefont {H.}~\bibnamefont {Jiang}},
  \bibinfo {author} {\bibfnamefont {Q.}~\bibnamefont {Gong}}, \ and\ \bibinfo
  {author} {\bibfnamefont {Y.}~\bibnamefont {Cheng}},\ }\href@noop {}
  {\bibfield  {journal} {\bibinfo  {journal} {Phys. Rev. Lett.}\ }\textbf
  {\bibinfo {volume} {116}},\ \bibinfo {pages} {143007} (\bibinfo {year}
  {2016})}\BibitemShut {NoStop}%
\bibitem [{\citenamefont {Azarm}\ \emph {et~al.}(2017)\citenamefont {Azarm},
  \citenamefont {Corkum},\ and\ \citenamefont {Polynkin}}]{Azarm}%
  \BibitemOpen
  \bibfield  {author} {\bibinfo {author} {\bibfnamefont {A.}~\bibnamefont
  {Azarm}}, \bibinfo {author} {\bibfnamefont {P.}~\bibnamefont {Corkum}}, \
  and\ \bibinfo {author} {\bibfnamefont {P.}~\bibnamefont {Polynkin}},\
  }\href@noop {} {\bibfield  {journal} {\bibinfo  {journal} {Phys. Rev. A}\
  }\textbf {\bibinfo {volume} {96}},\ \bibinfo {pages} {051401(R)} (\bibinfo
  {year} {2017})}\BibitemShut {NoStop}%
\bibitem [{\citenamefont {Arissian}\ \emph {et~al.}(2018)\citenamefont
  {Arissian}, \citenamefont {Kamer}, \citenamefont {Rastegari.A.},
  \citenamefont {Villeneuve},\ and\ \citenamefont {Diels}}]{Arissian18}%
  \BibitemOpen
  \bibfield  {author} {\bibinfo {author} {\bibfnamefont {L.}~\bibnamefont
  {Arissian}}, \bibinfo {author} {\bibfnamefont {B.}~\bibnamefont {Kamer}},
  \bibinfo {author} {\bibnamefont {Rastegari.A.}}, \bibinfo {author}
  {\bibfnamefont {D.}~\bibnamefont {Villeneuve}}, \ and\ \bibinfo {author}
  {\bibfnamefont {J.}~\bibnamefont {Diels}},\ }\href@noop {} {\bibfield
  {journal} {\bibinfo  {journal} {Phys. Rev. A}\ }\textbf {\bibinfo {volume}
  {98}},\ \bibinfo {pages} {053438} (\bibinfo {year} {2018})}\BibitemShut
  {NoStop}%
\bibitem [{\citenamefont {Britton}\ \emph {et~al.}(2018)\citenamefont
  {Britton}, \citenamefont {Laferri\`{e}re}, \citenamefont {Ko}, \citenamefont
  {Li}, \citenamefont {Kong}, \citenamefont {Brown}, \citenamefont {Naumov},
  \citenamefont {Zhang}, \citenamefont {Arissian},\ and\ \citenamefont
  {Corkum}}]{Britton18}%
  \BibitemOpen
  \bibfield  {author} {\bibinfo {author} {\bibfnamefont {M.}~\bibnamefont
  {Britton}}, \bibinfo {author} {\bibfnamefont {P.}~\bibnamefont
  {Laferri\`{e}re}}, \bibinfo {author} {\bibfnamefont {D.}~\bibnamefont {Ko}},
  \bibinfo {author} {\bibfnamefont {Z.}~\bibnamefont {Li}}, \bibinfo {author}
  {\bibfnamefont {F.}~\bibnamefont {Kong}}, \bibinfo {author} {\bibfnamefont
  {G.}~\bibnamefont {Brown}}, \bibinfo {author} {\bibfnamefont
  {A.}~\bibnamefont {Naumov}}, \bibinfo {author} {\bibfnamefont
  {C.}~\bibnamefont {Zhang}}, \bibinfo {author} {\bibfnamefont
  {L.}~\bibnamefont {Arissian}}, \ and\ \bibinfo {author} {\bibfnamefont
  {P.}~\bibnamefont {Corkum}},\ }\href@noop {} {\bibfield  {journal} {\bibinfo
  {journal} {Phys. Rev. Lett.}\ }\textbf {\bibinfo {volume} {120}},\ \bibinfo
  {pages} {133208} (\bibinfo {year} {2018})}\BibitemShut {NoStop}%
\bibitem [{\citenamefont {Britton}\ \emph {et~al.}(2019)\citenamefont
  {Britton}, \citenamefont {Lytova}, \citenamefont {Laferri\`{e}re},
  \citenamefont {Peng}, \citenamefont {Ko}, \citenamefont {Polynkin},
  \citenamefont {Villeneuve}, \citenamefont {Zhang}, \citenamefont {Spanner},
  \citenamefont {Arissian},\ and\ \citenamefont {Corkum}}]{Britton19}%
  \BibitemOpen
  \bibfield  {author} {\bibinfo {author} {\bibfnamefont {M.}~\bibnamefont
  {Britton}}, \bibinfo {author} {\bibfnamefont {M.}~\bibnamefont {Lytova}},
  \bibinfo {author} {\bibfnamefont {P.}~\bibnamefont {Laferri\`{e}re}},
  \bibinfo {author} {\bibfnamefont {P.}~\bibnamefont {Peng}}, \bibinfo {author}
  {\bibfnamefont {D.}~\bibnamefont {Ko}}, \bibinfo {author} {\bibfnamefont
  {P.}~\bibnamefont {Polynkin}}, \bibinfo {author} {\bibfnamefont
  {D.}~\bibnamefont {Villeneuve}}, \bibinfo {author} {\bibfnamefont
  {C.}~\bibnamefont {Zhang}}, \bibinfo {author} {\bibfnamefont
  {M.}~\bibnamefont {Spanner}}, \bibinfo {author} {\bibfnamefont
  {L.}~\bibnamefont {Arissian}}, \ and\ \bibinfo {author} {\bibfnamefont
  {P.}~\bibnamefont {Corkum}},\ }\href@noop {} {\bibfield  {journal} {\bibinfo
  {journal} {Phys. Rev. A}\ }\textbf {\bibinfo {volume} {100}},\ \bibinfo
  {pages} {013406} (\bibinfo {year} {2019})}\BibitemShut {NoStop}%
\bibitem [{\citenamefont {Kartashov}\ \emph {et~al.}()\citenamefont
  {Kartashov}, \citenamefont {Haessler}, \citenamefont {Ali\v{s}auskas},
  \citenamefont {Andriukaitis}, \citenamefont {Pug\v{z}lys}, \citenamefont
  {Baltu\v{s}ka}, \citenamefont {M\"{o}hring}, \citenamefont {Starukhin},
  \citenamefont {Motzkus}, \citenamefont {Zheltikov}, \citenamefont {Richter},
  \citenamefont {Morales}, \citenamefont {Smirnova}, \citenamefont {Ivanov},\
  and\ \citenamefont {Spanner}}]{Kartashov14}%
  \BibitemOpen
  \bibfield  {author} {\bibinfo {author} {\bibfnamefont {D.}~\bibnamefont
  {Kartashov}}, \bibinfo {author} {\bibfnamefont {S.}~\bibnamefont {Haessler}},
  \bibinfo {author} {\bibfnamefont {S.}~\bibnamefont {Ali\v{s}auskas}},
  \bibinfo {author} {\bibfnamefont {G.}~\bibnamefont {Andriukaitis}}, \bibinfo
  {author} {\bibfnamefont {A.}~\bibnamefont {Pug\v{z}lys}}, \bibinfo {author}
  {\bibfnamefont {A.}~\bibnamefont {Baltu\v{s}ka}}, \bibinfo {author}
  {\bibfnamefont {J.}~\bibnamefont {M\"{o}hring}}, \bibinfo {author}
  {\bibfnamefont {D.}~\bibnamefont {Starukhin}}, \bibinfo {author}
  {\bibfnamefont {M.}~\bibnamefont {Motzkus}}, \bibinfo {author} {\bibfnamefont
  {A.~M.}\ \bibnamefont {Zheltikov}}, \bibinfo {author} {\bibfnamefont
  {M.}~\bibnamefont {Richter}}, \bibinfo {author} {\bibfnamefont
  {F.}~\bibnamefont {Morales}}, \bibinfo {author} {\bibfnamefont
  {O.}~\bibnamefont {Smirnova}}, \bibinfo {author} {\bibfnamefont {M.~Y.}\
  \bibnamefont {Ivanov}}, \ and\ \bibinfo {author} {\bibfnamefont
  {M.}~\bibnamefont {Spanner}},\ }in\ \href {www.scopus.com} {\emph {\bibinfo
  {booktitle} {Research in Optical Sciences (Optical Society of America,
  Washington, 2014), p. HTh4B.5}}}\BibitemShut {NoStop}%
\bibitem [{\citenamefont {Mysyrowicz}\ \emph {et~al.}(2019)\citenamefont
  {Mysyrowicz}, \citenamefont {Danylo}, \citenamefont {Houard}, \citenamefont
  {Tikhonchuk}, \citenamefont {Zhang}, \citenamefont {Fan}, \citenamefont
  {Liang}, \citenamefont {Zhuang}, \citenamefont {Yuan},\ and\ \citenamefont
  {Liu}}]{Mysyrowicz}%
  \BibitemOpen
  \bibfield  {author} {\bibinfo {author} {\bibfnamefont {A.}~\bibnamefont
  {Mysyrowicz}}, \bibinfo {author} {\bibfnamefont {R.}~\bibnamefont {Danylo}},
  \bibinfo {author} {\bibfnamefont {A.}~\bibnamefont {Houard}}, \bibinfo
  {author} {\bibfnamefont {V.}~\bibnamefont {Tikhonchuk}}, \bibinfo {author}
  {\bibfnamefont {X.}~\bibnamefont {Zhang}}, \bibinfo {author} {\bibfnamefont
  {Z.}~\bibnamefont {Fan}}, \bibinfo {author} {\bibfnamefont {Q.}~\bibnamefont
  {Liang}}, \bibinfo {author} {\bibfnamefont {S.}~\bibnamefont {Zhuang}},
  \bibinfo {author} {\bibfnamefont {L.}~\bibnamefont {Yuan}}, \ and\ \bibinfo
  {author} {\bibfnamefont {Y.}~\bibnamefont {Liu}},\ }\href@noop {} {\bibfield
  {journal} {\bibinfo  {journal} {APL Photonics}\ }\textbf {\bibinfo {volume}
  {4}},\ \bibinfo {pages} {110807} (\bibinfo {year} {2019})}\BibitemShut
  {NoStop}%
\bibitem [{\citenamefont {Kocharovskaya}(1992)}]{OlgaK1}%
  \BibitemOpen
  \bibfield  {author} {\bibinfo {author} {\bibfnamefont {O.}~\bibnamefont
  {Kocharovskaya}},\ }\href@noop {} {\bibfield  {journal} {\bibinfo  {journal}
  {Phys. Rep.}\ }\textbf {\bibinfo {volume} {219}},\ \bibinfo {pages} {175}
  (\bibinfo {year} {1992})}\BibitemShut {NoStop}%
\bibitem [{\citenamefont {Kocharovskaya}\ and\ \citenamefont
  {Khanin}(1988)}]{OlgaK2}%
  \BibitemOpen
  \bibfield  {author} {\bibinfo {author} {\bibfnamefont {O.}~\bibnamefont
  {Kocharovskaya}}\ and\ \bibinfo {author} {\bibfnamefont {Y.}~\bibnamefont
  {Khanin}},\ }\href@noop {} {\bibfield  {journal} {\bibinfo  {journal} {Pis'ma
  Zh. Eksp. Teor. Fiz.}\ }\textbf {\bibinfo {volume} {48}},\ \bibinfo {pages}
  {581} (\bibinfo {year} {1988})},\ \bibinfo {note} {[JETP Lett. {\bf 48}, 630
  (1988)]}\BibitemShut {NoStop}%
\bibitem [{\citenamefont {Khanin}\ and\ \citenamefont
  {Kocharovskaya}(1990)}]{OlgaK3}%
  \BibitemOpen
  \bibfield  {author} {\bibinfo {author} {\bibfnamefont {Y.}~\bibnamefont
  {Khanin}}\ and\ \bibinfo {author} {\bibfnamefont {O.}~\bibnamefont
  {Kocharovskaya}},\ }\href@noop {} {\bibfield  {journal} {\bibinfo  {journal}
  {J. Opt. Soc. Am. B}\ }\textbf {\bibinfo {volume} {7}},\ \bibinfo {pages}
  {2016} (\bibinfo {year} {1990})}\BibitemShut {NoStop}%
\bibitem [{\citenamefont {Richter}\ \emph {et~al.}()\citenamefont {Richter},
  \citenamefont {Lytova}, \citenamefont {Morales}, \citenamefont {Haessler},
  \citenamefont {Smirnova}, \citenamefont {Spanner},\ and\ \citenamefont
  {Ivanov}}]{MBI_airlasing_paper}%
  \BibitemOpen
  \bibfield  {author} {\bibinfo {author} {\bibfnamefont {M.}~\bibnamefont
  {Richter}}, \bibinfo {author} {\bibfnamefont {M.}~\bibnamefont {Lytova}},
  \bibinfo {author} {\bibfnamefont {F.}~\bibnamefont {Morales}}, \bibinfo
  {author} {\bibfnamefont {S.}~\bibnamefont {Haessler}}, \bibinfo {author}
  {\bibfnamefont {O.}~\bibnamefont {Smirnova}}, \bibinfo {author}
  {\bibfnamefont {M.}~\bibnamefont {Spanner}}, \ and\ \bibinfo {author}
  {\bibfnamefont {M.}~\bibnamefont {Ivanov}},\ }\href@noop {} {\ }\bibinfo
  {note} {ArXiv:2001.08081 (submitted, 2020)}\BibitemShut {NoStop}%
\bibitem [{\citenamefont {Klynning}\ and\ \citenamefont
  {Pag\`es}(1982)}]{Klyn82}%
  \BibitemOpen
  \bibfield  {author} {\bibinfo {author} {\bibfnamefont {L.}~\bibnamefont
  {Klynning}}\ and\ \bibinfo {author} {\bibfnamefont {P.}~\bibnamefont
  {Pag\`es}},\ }\href {www.scopus.com} {\bibfield  {journal} {\bibinfo
  {journal} {Phys. Scr.}\ }\textbf {\bibinfo {volume} {25}},\ \bibinfo {pages}
  {543} (\bibinfo {year} {1982})}\BibitemShut {NoStop}%
\bibitem [{\citenamefont {Herzberg}(1989)}]{Herzberg}%
  \BibitemOpen
  \bibfield  {author} {\bibinfo {author} {\bibfnamefont {G.}~\bibnamefont
  {Herzberg}},\ }\href@noop {} {\emph {\bibinfo {title} {Molecular Spectra and
  Molecular Structure: Volume I - Spectra of Diatomic Molecules.}}},\ \bibinfo
  {edition} {2nd}\ ed.\ (\bibinfo  {publisher} {Krieger Publishing Company},\
  \bibinfo {address} {Florida USA},\ \bibinfo {year} {1989})\BibitemShut
  {NoStop}%
\bibitem [{\citenamefont {Boyd}(2008)}]{Boyd}%
  \BibitemOpen
  \bibfield  {author} {\bibinfo {author} {\bibfnamefont {R.}~\bibnamefont
  {Boyd}},\ }\href@noop {} {\emph {\bibinfo {title} {Nonlinear Optics}}},\
  \bibinfo {edition} {3rd}\ ed.\ (\bibinfo  {publisher} {Elsevier,Academis
  Press},\ \bibinfo {year} {2008})\BibitemShut {NoStop}%
\bibitem [{\citenamefont {Keldysh}(1964)}]{Keldysh}%
  \BibitemOpen
  \bibfield  {author} {\bibinfo {author} {\bibfnamefont {L.}~\bibnamefont
  {Keldysh}},\ }\href@noop {} {\bibfield  {journal} {\bibinfo  {journal} {Zh.
  Eksp. Teor. Fiz.}\ }\textbf {\bibinfo {volume} {47}},\ \bibinfo {pages}
  {1945} (\bibinfo {year} {1964})},\ \bibinfo {note} {[Sol. Phys. JETP Lett.
  {\bf 20}, 1307 (1965)]}\BibitemShut {NoStop}%
\bibitem [{\citenamefont {Pavi\v{c}i\'c}\ \emph {et~al.}(2007)\citenamefont
  {Pavi\v{c}i\'c}, \citenamefont {Lee}, \citenamefont {Rayner}, \citenamefont
  {Corkum},\ and\ \citenamefont {Villeneuve}}]{Pavicic}%
  \BibitemOpen
  \bibfield  {author} {\bibinfo {author} {\bibfnamefont {D.}~\bibnamefont
  {Pavi\v{c}i\'c}}, \bibinfo {author} {\bibfnamefont {K.}~\bibnamefont {Lee}},
  \bibinfo {author} {\bibfnamefont {D.}~\bibnamefont {Rayner}}, \bibinfo
  {author} {\bibfnamefont {P.}~\bibnamefont {Corkum}}, \ and\ \bibinfo {author}
  {\bibfnamefont {D.}~\bibnamefont {Villeneuve}},\ }\href@noop {} {\bibfield
  {journal} {\bibinfo  {journal} {Phys. Rev. Lett.}\ }\textbf {\bibinfo
  {volume} {98}},\ \bibinfo {pages} {243001(4)} (\bibinfo {year}
  {2007})}\BibitemShut {NoStop}%
\bibitem [{\citenamefont {Spanner}\ and\ \citenamefont
  {Patchkovskii}(2013)}]{Spanner13}%
  \BibitemOpen
  \bibfield  {author} {\bibinfo {author} {\bibfnamefont {M.}~\bibnamefont
  {Spanner}}\ and\ \bibinfo {author} {\bibfnamefont {S.}~\bibnamefont
  {Patchkovskii}},\ }\href@noop {} {\bibfield  {journal} {\bibinfo  {journal}
  {Chem. Phys.}\ }\textbf {\bibinfo {volume} {414}},\ \bibinfo {pages} {10}
  (\bibinfo {year} {2013})}\BibitemShut {NoStop}%
\bibitem [{\citenamefont {Schmidt}\ \emph {et~al.}(1993)\citenamefont
  {Schmidt}, \citenamefont {Baldridge}, \citenamefont {Boatz}, \citenamefont
  {Elbert}, \citenamefont {Gordon}, \citenamefont {Jensen}, \citenamefont
  {Koseki}, \citenamefont {Matsunaga}, \citenamefont {Nguyen}, \citenamefont
  {Su}, \citenamefont {Windus}, \citenamefont {Dupuis},\ and\ \citenamefont
  {Montgomery~Jr}}]{GAMESS}%
  \BibitemOpen
  \bibfield  {author} {\bibinfo {author} {\bibfnamefont {M.}~\bibnamefont
  {Schmidt}}, \bibinfo {author} {\bibfnamefont {K.}~\bibnamefont {Baldridge}},
  \bibinfo {author} {\bibfnamefont {J.}~\bibnamefont {Boatz}}, \bibinfo
  {author} {\bibfnamefont {S.}~\bibnamefont {Elbert}}, \bibinfo {author}
  {\bibfnamefont {M.}~\bibnamefont {Gordon}}, \bibinfo {author} {\bibfnamefont
  {J.}~\bibnamefont {Jensen}}, \bibinfo {author} {\bibfnamefont
  {S.}~\bibnamefont {Koseki}}, \bibinfo {author} {\bibfnamefont
  {N.}~\bibnamefont {Matsunaga}}, \bibinfo {author} {\bibfnamefont
  {K.}~\bibnamefont {Nguyen}}, \bibinfo {author} {\bibfnamefont
  {S.}~\bibnamefont {Su}}, \bibinfo {author} {\bibfnamefont {T.}~\bibnamefont
  {Windus}}, \bibinfo {author} {\bibfnamefont {M.}~\bibnamefont {Dupuis}}, \
  and\ \bibinfo {author} {\bibfnamefont {J.}~\bibnamefont {Montgomery~Jr}},\
  }\href@noop {} {\bibfield  {journal} {\bibinfo  {journal} {J. Comput. Chem.}\
  }\textbf {\bibinfo {volume} {14}},\ \bibinfo {pages} {1347} (\bibinfo {year}
  {1993})}\BibitemShut {NoStop}%
\bibitem [{\citenamefont {Bullough}\ \emph {et~al.}(1979)\citenamefont
  {Bullough}, \citenamefont {Jack}, \citenamefont {Kitchenside},\ and\
  \citenamefont {Saunders}}]{SolitonPaper}%
  \BibitemOpen
  \bibfield  {author} {\bibinfo {author} {\bibfnamefont {R.}~\bibnamefont
  {Bullough}}, \bibinfo {author} {\bibfnamefont {P.}~\bibnamefont {Jack}},
  \bibinfo {author} {\bibfnamefont {P.}~\bibnamefont {Kitchenside}}, \ and\
  \bibinfo {author} {\bibfnamefont {R.}~\bibnamefont {Saunders}},\ }\href@noop
  {} {\bibfield  {journal} {\bibinfo  {journal} {Physica Scripta}\ }\textbf
  {\bibinfo {volume} {20}},\ \bibinfo {pages} {364} (\bibinfo {year}
  {1979})}\BibitemShut {NoStop}%
\bibitem [{\citenamefont {Quarteroni}\ \emph {et~al.}(2007)\citenamefont
  {Quarteroni}, \citenamefont {Sacco},\ and\ \citenamefont {Saleri}}]{Quart}%
  \BibitemOpen
  \bibfield  {author} {\bibinfo {author} {\bibfnamefont {A.}~\bibnamefont
  {Quarteroni}}, \bibinfo {author} {\bibfnamefont {R.}~\bibnamefont {Sacco}}, \
  and\ \bibinfo {author} {\bibfnamefont {F.}~\bibnamefont {Saleri}},\
  }\href@noop {} {\emph {\bibinfo {title} {Numerical Mathematics}}},\ \bibinfo
  {edition} {2nd}\ ed.\ (\bibinfo  {publisher} {Springer-Verlag Berlin
  Heidelberg},\ \bibinfo {year} {2007})\BibitemShut {NoStop}%
\bibitem [{\citenamefont {Strikwerda}(2004)}]{Strik}%
  \BibitemOpen
  \bibfield  {author} {\bibinfo {author} {\bibfnamefont {J.}~\bibnamefont
  {Strikwerda}},\ }\href@noop {} {\emph {\bibinfo {title} {Finite Difference
  Schemes and Partial Differential Equations.}}},\ \bibinfo {edition} {2nd}\
  ed.\ (\bibinfo  {publisher} {Society for Industrial and Applied Mathematics
  (SIAM)},\ \bibinfo {address} {Philadelphia, PA},\ \bibinfo {year} {2004})\
  pp.\ \bibinfo {pages} {xii+435}\BibitemShut {NoStop}%
\end{thebibliography}

\providecommand{\noopsort}[1]{}\providecommand{\singleletter}[1]{#1}%

\end{document}